\RequirePackage[2020-02-02]{latexrelease} % Without the package specificaiton, the error message occure. 
\documentclass[times, twoside]{zHenriquesLab-StyleBioRxiv}

\leadauthor{Lead Author} 

\usepackage{soul}
\usepackage{etoolbox}

\usepackage{ushort, mathtools}
\usepackage{widetext}
\makeatletter
\newcommand*{\newbibstartnumber}[1]{%
  \apptocmd{\thebibliography}{%
    \global\c@NAT@ctr #1\relax
    \addtocounter{NAT@ctr}{-1}%
  }{}{}%
}
\makeatother

\begin{document}
\title{B\'ezier interpolation improves the inference of dynamical models from data}
\shorttitle{Short title}

%Probability-preserving interpolation improves the inference of dynamical models from data (???)
%Probability-preserving non-linear interpolation improves the inference of dynamical models from data (???)

\author[1]{Kai Shimagaki}
\author[1,2,$\dagger$]{John P.~Barton}

%ORCiDs
% XYZ: 
% JPB: 0000-0003-1467-421X

\affil[1]{Department of Physics and Astronomy, University of California, Riverside, USA.
\textsuperscript{2}Department of Computational and Systems Biology, University of Pittsburgh School of Medicine, USA.
%*These authors contributed equally to this work. 
$\dagger$Address correspondence to: jpbarton@pitt.edu.}

\maketitle

%TC:break Abstract
%the command above serves to have a word count for the abstract
\begin{abstract}
\noindent
Many dynamical systems, from quantum many-body systems to evolving populations to financial markets, 
% examples -- these are other examples of stochastic processes in standard physics (I think)
% 1). active matter physics\cite{bonilla2019active, flenner2020active}
% 2). quantum many-body systems\cite{weidinger2017floquet, oka2018floquet, jung1993periodically}
are described by stochastic processes. 
Parameters characterizing such processes can often be inferred using information integrated over stochastic paths.
However, estimating time-integrated quantities from real data with limited time resolution is challenging. 
Here, we propose a framework for accurately estimating time-integrated quantities using B\'ezier interpolation. 
We applied our approach to two dynamical inference problems: determining fitness parameters for evolving populations and inferring forces driving Ornstein-Uhlenbeck processes. 
We found that B\'ezier interpolation reduces the estimation bias for both dynamical inference problems. This improvement was especially noticeable for data sets with limited time resolution.
Our method could be broadly applied to improve accuracy for other dynamical inference problems using finitely sampled data.
\end{abstract}

\subsection*{Introduction}\phantomsection\label{sec:introduction}
Stochastic processes are ubiquitous in nature. 
In biology, the evolution of genetic sequences can be formulated as a stochastic process.
The Wright-Fisher (WF) model\cite{ewens2004mathematical}, a discrete-time stochastic process, has been used to study the evolution of organisms from viruses\cite{foll2014influenza, ferrer2016approximate, sohail2021mpl} to humans\cite{mathieson2013estimating}.
Models such as the Ornstein-Uhlenbeck (OU) process\cite{iacus2008simulation, gillespie1996exact} have been applied to describe a wide range of phenomena, from the fluctuation of currency exchange rates\cite{roberts2004bayesian} and cell migration\cite{dieterich2008anomalous} to driven quantum many-body systems\cite{jung1993periodically}.

Appropriate model parameters are needed to accurately describe the behavior or real systems.
To infer such parameters from data, it is often necessary to compute statistics over a \emph{path}, i.e., a complete realization of the stochastic processes. 
%For example, an accumulation of stochastic variables over the path gives its estimators, parameter characterizes the restoring force in the OU model\cite{liptser1977statistics}.
For example, the restoring force of the OU process can be estimated by taking the ratio of the deviation from the equilibrium position and the magnitude of the intrinsic fluctuations, both integrated over a stochastic path\cite{phillips2009maximum, liptser1977statistics}.

However, real data often consists of incomplete, occasional measurements of a system, which may also be limited by experimental constraints. 
This makes it more difficult to accurately estimate model parameters since statistics over the path must be estimated from incomplete information. 
A workaround used in a previous study\cite{sohail2021mpl} for this problem is to use linear interpolation to estimate the state of the system between the observed data points.
%Linear interpolation has a favorable property that any interpolation point of probability densitity can be also considered as probability density. -- suggest just deleting since this is tangential
However, this approximation may fail when gaps in time are large enough such that the behavior of the system is highly nonlinear\cite{wittman2005mathematical}. 
% However, this is a rough approximation that may fail when gaps in time are large enough such that the behavior of system is strongly nonlinear. -- suggest replacing the sentence above with something like this that is less specific and gives us room to explain later
%Even though many stochastic processes evolve nonlinearly, few interpolation frameworks can fully count nonlinearity and are not applicable to probability densities. 
%Related interpolation methods for nonlinear systems require an explicit form of the underlying nonlinear function and/or its analytical solution for stochastic processes\cite{meditch1973survey}. 
%Lastly, there are few studies that quantitatively explain in which circumstances linear interpolation is inadequate and nonlinear interpolation becomes efficient\cite{walia2006analysis}.

Here, we propose a tractable nonlinear interpolation framework using B\'ezier curves. 
In addition to incorporating nonlinearity, this approach has the added advantage of conserving sums of categorical variables, which is not guaranteed under arbitrary nonlinear transformations of data.
This property can be especially useful for conserved quantities such as probabilities.
%in order to overcome the limitations: few existing interpolation frameworks can account for nonlinearity and apply it to probability densities, and few studies have quantitatively explained when nonlinear interpolation is necessary. 
Historically, the B{\'e}zier method has been used in computer graphics to draw smooth curves\cite{farouki2012bernstein, choi2008path, simba2014real, forrest1972interactive}.

We applied B{\'e}zier interpolation to two example problems: inferring natural selection in evolving populations through the WF model and inferring restoring forces for OU processes. 
Here, our method reduces estimation bias and improves the precision of model inferences.
Furthermore, we show that the autocorrelation function of statistics over a path identifies time scales over which nonlinear interpolation is particularly effective, which is consistent with our observations in simulations.
We show that B{\'e}zier interpolation can generically improve solutions of dynamical inference problems by accurately estimating statistics over stochastic paths.
We expect that this nonlinear interpolation method can improve a wide range of dynamical inference problems beyond the specific examples we consider, such as parameter estimation for stochastic differential equations. 
Our approach is particularly well-suited for situations in which difficult to obtain samples with good time resolution.

\subsection*{B\'ezier interpolation}\phantomsection\label{sec:Bezier_definition}
Consider a function $x(t)$ sampled at discrete times $t_k$ for $k\in\{0, 1, \ldots, K\}$. 
Then the interpolated value of the function $x^{(k)}_B(t)$ between two successive discrete time points $t_k$ and $t_{k+1}$ is given by
\begin{equation}
 x^{(k)}_B(t) = \sum_{n=0}^P \beta_n\left(\frac{t-t_k}{t_{k+1}-t_k}\right) \phi_{n}^{(k)}((x(t_{k'}))_{k'=0}^K)\,. \label{eq:Bezier_interpolation_simple}
\end{equation}
% B\'ezier interpolation is a continuous function depending on continuous polynomial functions and can connect whole data points seamlessly and smoothly. 
% Any interpolation point at an arbitrary time interval between a discrete time $t_k$ and its successive time $t_{k+1}$ for $k\in \{0,\ldots, K\}$ can be represented as a B\'ezier interpolation $B^{(k,k+1)}(\tau)$, depending on a continuous variable $\tau\in[0,1]$ , 
Here, $\beta_n$ is the $n$th Bernstein basis polynomial of degree $P$, with $\beta_n(\tau) = {P \choose n}\tau^n (1-\tau)^{P-n} \geq 0$. 
The control points $\phi_n^{(k)}((x(t_{k'}))_{k'=0}^K)$ depend on the ensemble of data points $(x(t_k))_{k=0}^K$ and determine the outline of the interpolation curves.

%B\'ezier curves interpolate between discrete data points through Bernstein polynomials.
%The Bernstein polynomial of degree $P$ has a series of bases, $\beta_n(\tau) = {P \choose n}\tau^n (1-\tau)^{P-n} \geq 0,~$ where $\tau \in [0,1]$, and $n\in\{0,1, \ldots, P\}$. 
%Here for simplicity we will consider cubic ($P=3$) interpolation, but this approach can be extended to polynomials of different degrees.
%Let $\left(x(t_k)\right)_{k=0}^{K}$ be the observed discrete data points.
%Cubic interpolation in a time interval $[t_k, t_{k+1}]$ then follows 
%\begin{equation}
% B^{(k, k+1)}(\tau) = \sum_{n=0}^3 \beta_n(\tau) %\phi_{n}^{(k)}(\{x(t_{k'})\})\,. \label{eq:Bezier_interpolation_simple}
%\end{equation}
%\textcolor{red}{We need to explain how this equation works. What is $\tau$? What is $B$? What are the $\phi$? What is $k^\prime$? What is $\boldsymbol{x}$? Also, what do the curly braces $\{\}$ around the $x$ values and the intervals below represent? The remainder of the section can be edited when the explanation is complete.}
% and the linear interpolation, i.e., Bernstein polynomial for $P=1$, defined in between these points is defined as follows,
% \begin{equation}
%  x^{k,k+1} (\tau) = \beta_0(\tau)x(t_k) + \beta_1(\tau) x(t_{k+1}) ~. \label{eq:linear_interpolation_simple}
% \end{equation}

%Similarly, we define the cubic, $P=3$ B\'ezier interpolation on some interval 

\begin{figure}%[tbhp]
\centering
\includegraphics[width=\linewidth]{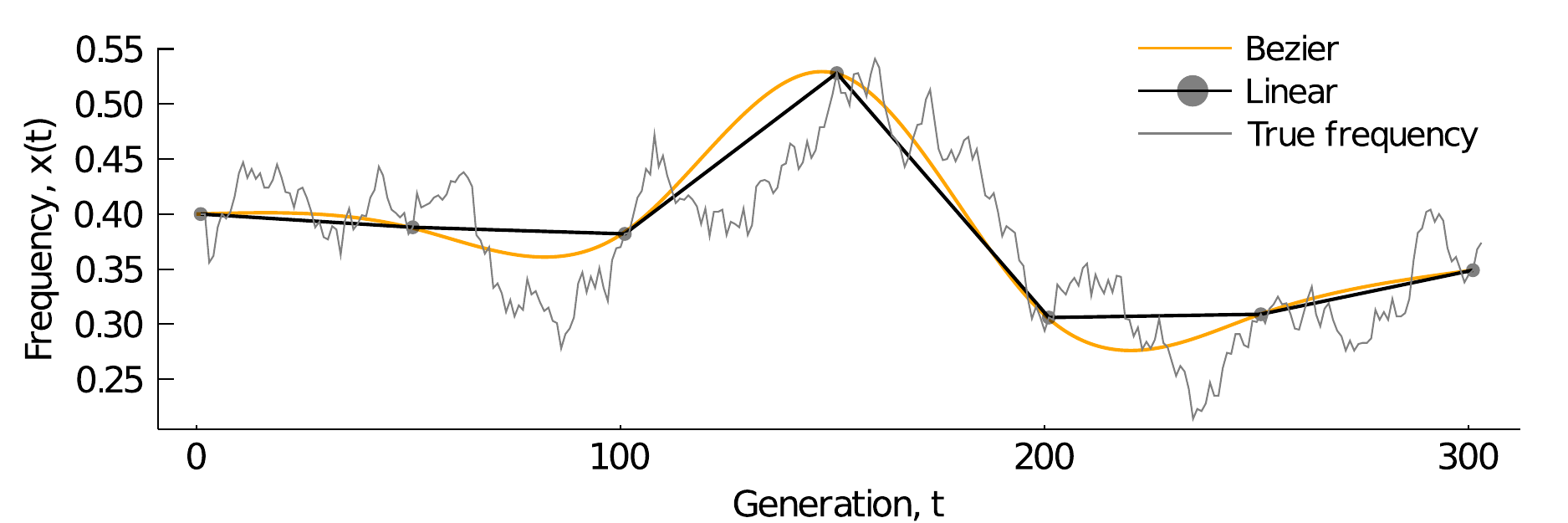}
\caption{ 
\textbf{B{\'e}zier interpolation generates smooth curves.} Cubic B\'ezier curves smoothly interpolate between discretely-sampled frequency trajectories generated from a Wright-Fisher model. 
\textit{Simulation parameters}.
$L=50$ sites, population size $N=10^3$, mutation rate $\mu = 10^{-3}$, with simulations over $T=300$ generations. 
Data points are sampled every 50 generations and interpolated using cubic B\'ezier and linear interpolation.
% This info is alread in the legend and doesn't need to be repeated -- The horizontal axis is time corresponding to each generation. The black and orange lines are linear- and B\'ezier- interpolation using a set of observed points with the sampling interval $\Delta t = 50$.
\label{fig:fig1_interpolations}}
\end{figure}

For simplicity we consider cubic ($P=3$) interpolation, but our approach can be extended to polynomials of different degrees $P$.
We impose the following conditions to ensure that the segment at each interval $[t_k, t_{k+1}]~ \forall k$ is seamlessly connected,
\begin{equation}
 \phi_{0}^{(k)}((x(t_{k'}))_{k'=0}^K) =x(t_k),~~~\phi_{3}^{(k)}((x(t_{k'}))_{k'=0}^K) =x(t_{k+1}) ~. \nonumber
\end{equation}
Other internal points $\{ ( \phi_{1}^{(k)}, \phi_{2}^{(k)} ) \}_{k=0}^{K-1} $, are obtained by solving an optimization problem that reflects continuity and smoothness constraints imposed on the curves (see \hyperref[{sec:Bezier_interpolation}]{Methods}, \textbf{Fig.~\ref{fig:fig1_interpolations}}). 
%B\'ezier interpolation realizes a smooth curve while keeping similar properties of linear interpolation (\textbf{Fig.~\ref{fig:fig1_interpolations}}). 
%We also found that almost the same result can be obtained by assuming the discrete data points are continuously concatenated and the overall length of the path is minimized (Methods). 
%The following discussion examines the most commonly used cubic B\'ezier curves for probability interpolation, however the following discussions hold for other orders of B\'ezier method.

% The Bernstein basis has some suitable properties that can interpolate probability densities\cite{bernstein1912demo, farouki2012bernstein}. A sum of the bases over the index is one, $\sum_n^P \beta_n(\tau)= 1$. An statistics integrated over the interpolation parameter $\tau$ yields a constant value, $\int_0^1 \beta_n(\tau) d \tau =1/(P+1)$. Moreover, the integral of a pair of bases $\int_0^1 \beta_n(\tau)\beta_m(\tau) d\tau$ has a simple closed form depending only on $n,m,$ and $P$. In fact, linear interpolation is a special case of Bernstein polynomial. 
% Thanks to these properties, computations for statistics integrated  over time can be replaced by fast algebraic operations.

\subsection*{Results}\phantomsection\label{sec:results}
To test the performance of B\'ezier interpolation in dynamical inference problems, we studied two stochastic models.
First, we consider the Wright-Fisher (WF) model, a fundamental mathematical model in biology used to study evolving populations.
Second, we examine the Ornstein-Uhlenbeck (OU) model, a simple stochastic process with wide-ranging applications in multiple disciplines.
Below, we introduce each model, show how model parameters can be estimated from stochastic paths, and consider how B\'ezier interpolation aids inference from finitely sampled data.
We also demonstrate conditions under which nonlinear interpolation is most useful for inference.

\subsection*{Wright-Fisher model of evolution}\phantomsection\label{sec:wf_model}
The WF model\cite{ewens2004mathematical} is a classical model in evolutionary biology. 
In this model, a population of $N$ individuals evolves over discrete generations under the influence of random mutations and natural selection.
Each individual is represented by a genetic sequence of length $L$.
For simplicity, we assume that each site in the genetic sequence is occupied by a mutant ($1$) or wild-type ($0$) nucleotide.
There are thus $M=2^L$ possible \textit{genotypes} (i.e., genetic sequences) in the population.

The state of the population is described by a genotype frequency vector $\boldsymbol{z}(t) = \left(z_a(t)\right)_{a=1}^M$, where $z_a(t)$ represents the frequency of individuals with genotype $a$ in the population at time $t$.
Frequencies are normalized such that $\sum_a z_a(t) = 1$.
Then, in the WF model, the probability of obtaining a genotype frequency vector $\boldsymbol{z}^\prime$ in the next generation is multinomial,
\begin{equation}
	p(\boldsymbol{z}'|\boldsymbol{z}(t)) = N! \prod_{a=1}^M \frac{{p_a\left(\boldsymbol{z}(t)\right)}^{Nz^\prime_a}}{(Nz^\prime_a)!}\,.  \nonumber \label{eq:WF_def} 
\end{equation}
Here $p_a(\boldsymbol{z}(t))$ is the effective succession probability of genotype $a$ due to natural selection and mutation, 
\begin{equation}\label{eq:effective_succession_prob_WF}
	p_a(\boldsymbol{z}(t)) \propto f_a z_a(t) + \sum_{b|b\neq a}( \mu_{ba}z_b(t)f_b - \mu_{ab}z_a(t)f_a)\,. 
\end{equation}
In (\ref{eq:effective_succession_prob_WF}), $f_a$ denotes the \textit{fitness} of genotype $a$.
Individuals with higher fitness values reproduce more readily than those with lower fitness values.
Here $\mu_{ab}$ is the probability to mutate from genotype $a$ to genotype $b$.

In principle, fitness values can be estimated from genetic sequence data by identifying the $f_a$ that are most likely to generate the observed evolutionary history of a population.
Given the enormous size of the genotype space, however, simplifying assumptions are often needed.
A common choice is to assume that fitness values are additive,
$f_a = 1+\sum_{i=1}^L \sigma_i^a s_i$, where $\sigma_i^a=1$ if the nucleotide at site $i$ in genotype $a$ is a mutant and $0$ otherwise. 
The $s_i$ are referred to as \textit{selection coefficients}, which are positive if the mutation at site $i$ is beneficial for reproduction and negative if mutation at site $i$ is deleterious.
Similarly, the mutation rate $\mu_{ab}$ can be simplified to a constant $\mu$ if genotypes $a$ and $b$ differ from one another by only a single mutation and zero otherwise.

Sohail et al.~solved this problem analytically in the limit that the population size $N\rightarrow\infty$ while the selection coefficients $s_i$ and mutation rate $\mu$ scale as $1/N$ (ref.~\cite{sohail2021mpl}).
In this case, the maximum \textit{a posteriori} vector of selection coefficients $\hat{\boldsymbol{s}} = \left(\hat{s}_i\right)_{i=1}^L$ that best explain the data are given by
\begin{align}\begin{aligned} \label{eq:sMAP}
    \hat{\boldsymbol{s}} &= \left(\int_{t_0}^{t_K}\!\!\!\!dt\,\mathrm{C}(t) + \gamma \mathrm{I} \right)^{-1} \\
    &\qquad\times\left[\boldsymbol{x}(t_K) - \boldsymbol{x}(t_0) - \mu\int_{t_0}^{t_K}\!\!\!\!dt\left(1 - 2\boldsymbol{x}(t)\right)
     \right]\,,
\end{aligned}\end{align}
where the time of observation runs from $t_0$ to $t_K$.
In \eqref{eq:sMAP}, $\boldsymbol{x}(t) = \left(x_i(t)\right)_{i=1}^{L}$ is a vector of mutant frequencies (i.e., the number of individuals in the population with a mutation at site $i$ at time $t$), and $\mathrm{C}(t)$ is the covariance matrix of mutant frequencies at time $t$.
Here $\gamma$ is the precision of a Gaussian prior distribution for the selection coefficients with mean zero, and $\mathrm{I}$ is the identity matrix.

Extensive past work has also considered numerical solutions to this problem\cite{ferrer2016approximate, lacerda2014population, tataru2017statistical, schraiber2016bayesian, foll2015wfabc, mathieson2013estimating, iranmehr2017clear}, though the analytical formula in \eqref{eq:sMAP} typically outperforms numerical approaches\cite{sohail2021mpl}.
Sohail et al.~referred to \eqref{eq:sMAP} as the marginal path likelihood (MPL) estimate for the selection coefficients, obtained by maximizing the posterior probability of an evolutionary history (i.e., a stochastic path) with respect to the selection coefficients.
The MPL approach has also been extended to consider more complex evolutionary models\cite{sohail2022inferring} and epidemiological dynamics\cite{lee2022inferring}.

\subsection*{B\'ezier interpolation for WF model inference}\phantomsection\label{sec:wf_bezier}
In practice, \eqref{eq:sMAP} is not straightforward to evaluate because data is not available in continuous time.
Instead, sequence data comes at discrete times $\left(t_k\right)_{k=0}^{K}$, which may also be spaced heterogeneously in time.
To solve this problem, we apply B\'ezier interpolation to finitely sampled mutant frequency trajectories. 
This allows us to analytically integrate both mutant frequency trajectories $\boldsymbol{x}(t)$ and covariances $\mathrm{C}(t)$, obtained by interpolating frequencies and computing $C_{ij}(t) = x_{ij}(t) - x_i(t) x_j(t)$.
Here $x_{ij}(t)$ is the pairwise frequency of individuals in the population at time $t$ that have mutations at both sites $i$ and $j$.

%However, estimating the covariance information requires careful consideration because the pairwise correlation signal between alleles is weaker than the signal from individual allele frequencies in general. 
%That means that datasets with short observation time intervals are preferred for estimating genetic fluctuation.

% Here we apply the B\'ezier interpolation method for the allele frequency trajectories and indirectly interpolate the covariance matrix.
% \begin{equation}
%  \Delta \hat{C}^{(k)} = \Delta t_k\int_{0}^1 
%  \left( 
%  \hat{B}^{(k, k+1)}(\tau) - \boldsymbol{B}^{(k, k+1)}(\tau) (\boldsymbol{B}^{(k, k+1)}(\tau))^T 
%  \right) \label{eq:ICov_Bezier}~.
% \end{equation}
% Where $\boldsymbol{B}^{(k, k+1)}(\tau)$ and $\hat{B}^{(k, k+1)}(\tau)$ are a column vector and matrix of B\'ezier interpolation defined in a time interval $[t_k, t_{k+1}]$ based on the definition  \eqref{eq:Bezier_interpolation_simple}.
% In the following sections, we apply the B\'ezier interpolation for statistics estimated over the evolution including covariance to improve the MPL prediction for selection coefficients, the additive fitness parameters of the WF model. 

%\subsection*{Inference of selection for the Wright-Fisher model}\phantomsection\label{subsec:application_WF_model}

\begin{figure}%[tbhp]
\centering
\includegraphics[width=\linewidth]{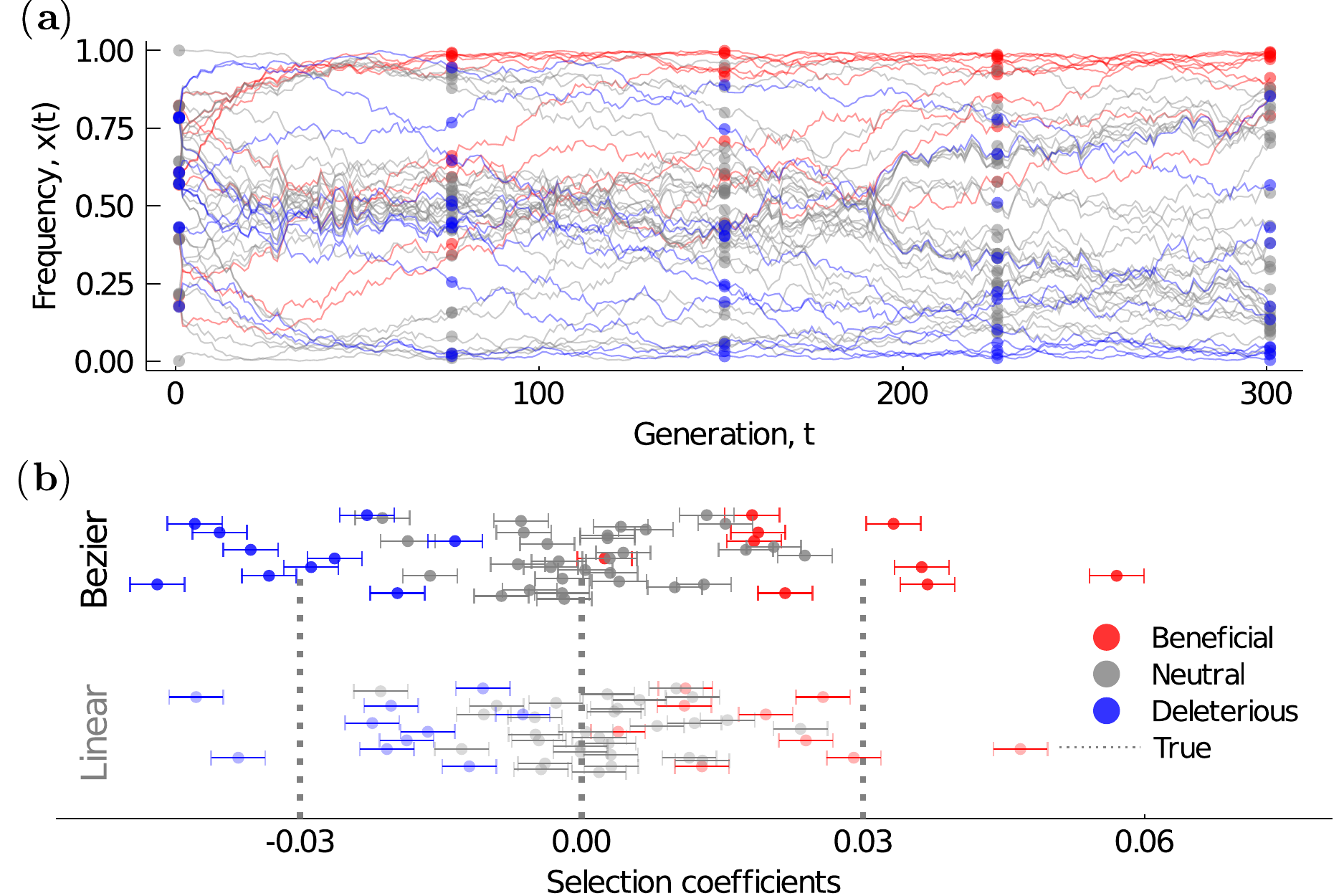}
\caption{
\textbf{B\'ezier interpolation reduces bias in estimated selection coefficients.}
%\textcolor{red}{For clarity, let's move the (a) (b) labels to the upper left corner of the panel that they describe. 
%Let's also extend the label for selection coefficients to include "coefficients" and the variable name.}
%\textcolor{red}{
%Many axis labels are too short to be clear, but almost never is an axis label too long!}
%\textcolor{blue}{Sorry, it seems there is no way to extend tick length in Julia. }
\textbf{(a)} Wright-Fisher simulation with selection and mutation. Each trajectory drawn as a solid line is true complete data, and filled circles are a subset of the complete data, which is observed every $\Delta t=75$ generations, and used for selection coefficient prediction.
\textbf{(b)} Selection coefficients for the frequency trajectories in (a) were estimated by MPL with B{\'e}zier and linear interpolation. Error bars are analytical standard deviations of the estimated selection coefficients, computed as the inverse of the square root of the diagonal entries of $\int_0^{t_K} dt \mathrm{C}(t)$.  MPL with B{\'e}zier interpolation greatly reduces estimation bias for inferred selection coefficients when the time interval between sampled observations is large. 
%\textcolor{red}{Question: are the mean values also from 100 simulations or just the error bars? Are the error bars aggregated across identical sites or computed for each site individually?}\textcolor{blue}{The “mean values” shown as filled circles correspond to estimated selection coefficients using trajectories generated from a single WF simulation. The WF simulation is randomly chosen among 100 simulations. Also, the error bars are computed for each site individually.
%}
%\textcolor{red}{Note on figure: some text begins with capital letters and some is in lower case -- it would be best to standardize this and have every phrase begin with either capital or lower case. Capital is standard for some journals (e.g., Nature family). For clarity, could beneficial/neutral/deleterious labels be moved further to the bottom right? We also need to label what the dashed line means.} \textcolor{blue}{Ok, these are done.}
\textit{Simulation parameters}.
$L=50$ sites with 10 beneficial, 10 deleterious, and 30 neutral mutations with selection coefficients of $s=0.03$, $s=-0.03$, and $s=0$, respectively. 
Other parameters of the WF models are the same as in \textbf{Fig.~\ref{fig:fig1_interpolations}}. 
%\textcolor{red}{[To prevent confusion if things get rearranged later, always use dynamical labels when referring to figures, equations, tables, etc.]}
\label{fig:fig3_selections_histogram}}
\end{figure}

To assess the performance of B\'ezier interpolation for inferring selection in the WF model, we generated a test data set by running 100 replicate simulations of WF evolution with identical parameters (\textbf{Fig.~\ref{fig:fig3_selections_histogram}a}). 
We then inferred selection coefficients from this data using MPL with linear and B\'ezier interpolation, applied to data sampled at discrete intervals $\Delta t = 75$ generations apart. 
While MPL with linear interpolation readily distinguishes between beneficial, neutral, and deleterious parameters, the inferred selection coefficients are shrunk towards zero. %\textcolor{blue}{Question: this might be miner, but does the phrase "MPL using linear interpolation make a clear distinction ..." sound contradict the following description of PPV? The following described that the precision of MPL using linear interpolation could be further improved by using Bezier interpolation in the sense of the PPV curves.}
However, parameters inferred using B\'ezier interpolation are distributed around their true values. (\textbf{Fig.~\ref{fig:fig3_selections_histogram}b}). 
B\'ezier interpolation reduces estimation bias due to long intervals between observations intervals by producing better estimates of underlying covariances (which we will quantify below).
%\textcolor{red}{[This seems like it must be the case, but is this observed in the data? Yunxiao has checked this when studying a different problem and might be a good person to consult with for how to plot/measure this} \textcolor{blue}{The associated result with this question is shown in Fig.5, the interpolation error of the integrated covariance matrix. In this sense, the Bezier interpolation can improve the estimation of underlying covariance matrices.} 
Here we used a regularization strength of $\gamma=0.1$, but similar results are obtained with different choices for the regularization (\hyperref[sec:methods]{Methods}).

\begin{figure}%[tbhp]
\centering
\includegraphics[width=\linewidth]{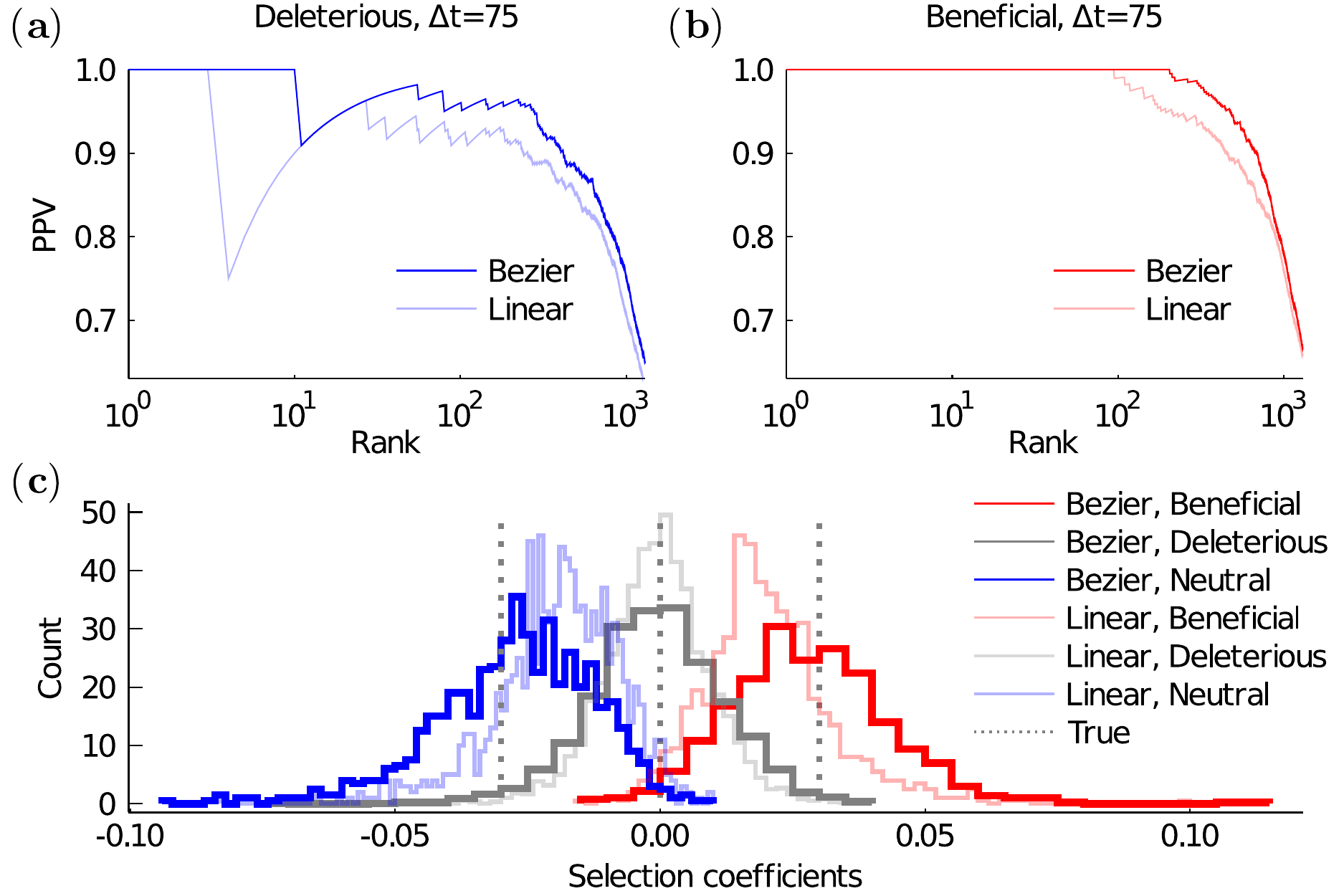}
\caption{
%\textcolor{red}{See related comments in previous figure. Typo: ``deliterious'' --> ``deleterious''. We should have just one bolded ``title'' sentence for each figure.} \textcolor{blue}{Ok, these are done.}
\textbf{MPL with B\'ezier interpolation improves prediction precision and reduces estimation bias.}
\textbf{(a)} Positive predictive value (PPV) curves, which quantify prediction precision for deleterious selection coefficients. When the observation time interval is longer ($\Delta t = 75$), the PPV curve for B\'ezier interpolation is universally higher than the curve for linear interpolation. \textbf{(b)} PPV curves for beneficial selection coefficients.
\textbf{(c)} The selection coefficient distributions estimated by MPL with linear interpolation visibly shrank toward zero and were biased, while distributions estimated by MPL with B\'ezeir interpolation did not considerably shrink and have the mean values near the true selection values.
\label{fig:fig4_PPV_Bezier_Linear}}
\end{figure}

Next we studied how B\'ezier interpolation affects our ability to classify mutations as beneficial or deleterious, which we evaluated by ranking mutations according to their inferred selection coefficients. 
This metric is distinct from the issue of biased estimation of selection coefficients.
We quantified classification accuracy using positive predictive value (PPV), ${\rm PPV}={\rm TP}/({\rm TP}+{\rm FP})$~, where ${\rm TP}$ and ${\rm FP}$ are the numbers of true positive and false positive predictions. 
%By sorting selection coefficients in descending (ascending) order and counting the number of actual beneficial and deleterious selection coefficients, we obtained PPV curves as a function of the rank. 
The PPV curves for beneficial/deleterious mutations estimated by MPL with B\'ezier interpolation are higher than those with linear interpolation, indicating more accurate classification (\textbf{Fig.~\ref{fig:fig4_PPV_Bezier_Linear}a-b}).
This can be understood by observing reduced overlap between the distribution of inferred selection coefficients for beneficial, neutral, and deleterious mutations using B\'ezier interpolation (\textbf{Fig.~\ref{fig:fig4_PPV_Bezier_Linear}c}).
%\textcolor{red}{[Maybe reverse the top and bottom for this figure?]} %\textcolor{blue}{Ok, it's done.}

% The B\'ezier method improves the prediction precision and biasness, which are qualitatively different: by uniformly scaling selection coefficients so that the mean value of the distribution matches the true selection values, the rank of the section coefficients remains identical, and the PPV curve does not change under the rescaling even if the bias is removed.

%MPL predicts selection accurately from limited observables, however the longer the sampling interval $\Delta t$ of the input data, the less accurate the inference, as reported below. Therefore, we will examine the effect of the $\Delta t$ on inference performance in detail (\textcolor{blue}{hyperlink: Section of Langevin and ACF and Supplement, $\Delta t$ dependency on PPV}). 
%The above discussions are not only for the binary variables case but can extend to more general multivariable cases, and we report the effect of B\'ezier interpolation for multivariable cases in the following section. 

\subsection*{Performance of B\'ezier interpolation on real data}\phantomsection\label{subsec:application_HIV}
To apply B\'ezier interpolation to biological sequence data, we extended the approach described above from binary variables to multivariates. 
This is necessary because DNA or RNA sequences have five possible states at each site, including four nucleotides and a ``gap'' symbol, which represents the absence of a nucleotide at a site that is present in other related sequences.

\begin{figure}%[tbhp]
\centering
\includegraphics[width=\linewidth]{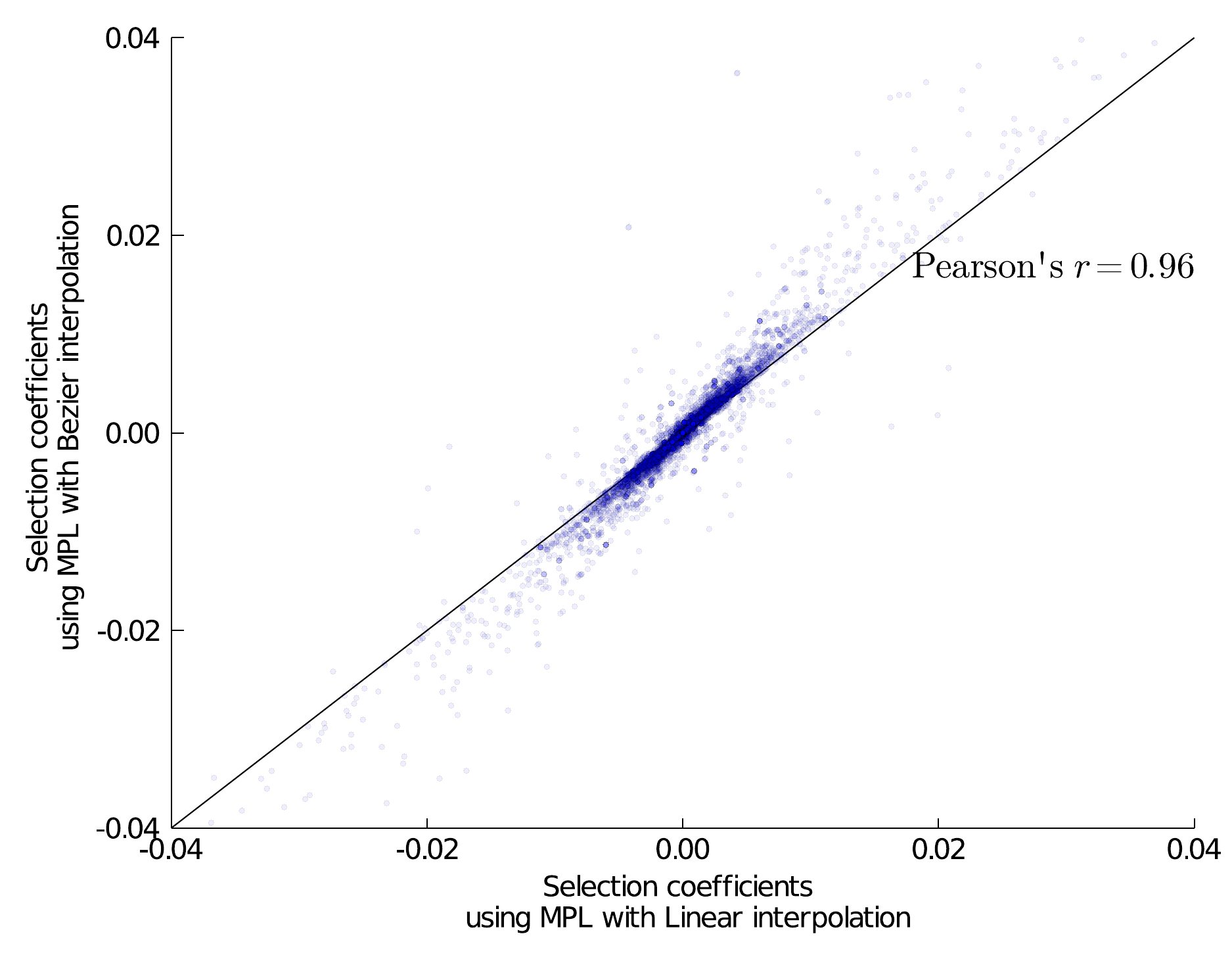}
\caption{ 
\textbf{HIV-1 selection coefficients estimated by MPL with B\'ezier interpolation are strongly correlated with those estimated with linear interpolation.}
%\textcolor{red}{-- what does this mean? Are we looking at selection coefficients for Env mutations across 13 individuals? Specifically Env or actually the $3^\prime$ end of the sequence? I'm not sure what vertical means in this context. The next sentence can be omitted because it is already shown in the plot. For clarity, let's call this ``Pearson's $r$'' in the plot instead of just ``Pearson''. Question on plot: it looks like the points are in different colors; what do the colors mean? If there is no (important) meaning, colors should probably all be the same. Dot outlines can also be removed to help show density more clearly}.  \textcolor{blue}{The different color meanings in the previous plot corresponded to different patients, but are not particularly important in this manuscript. I modified the figure and its caption considering the other suggestions.}
Consistent with simulation results in \textbf{Fig.~\ref{fig:fig3_selections_histogram}}, B\'ezier interpolation typically yields larger estimated selection coefficients. 
Selection coefficients were obtained for roughly 50 to 900 mutations per individual and sequencing region.
Samples were obtained from 3-9 times per individual, with 7-40 sequences per time point. 
Sequences were collected frequently early in infection with $\Delta t \sim 10$ days, stretching to 100-200 days late in infection.
Mutation rates from past studies\cite{zanini2015population} were used to estimate selection coefficients. 
The regularization strength is $\gamma = 10$ in both linear and B\'ezier cases. 
%For the real data, where each site is occupied by one of the $q=5$ states, it is possible for eigenvalues of the integrated covariance matrix to become negative. 
%For the actual data, each site is occupied by one of the $q=5$ states, and samples are taken more frequently earlier, with longer sampling intervals later. 
%Eigenvalues of the integrated covariance matrix can become negative due to scarce and heterogeneous sampling of sequences, and interpolation becomes unstable.
%To prevent this, we introduced mean frequency trajectory points at their midpoints for sampling intervals greater than 50 days.
%This is because, while the sum of all $q$ frequencies at a site is always equal to one, individual frequencies can occasionally become $<0$ or $>1$. To prevent this, we adjusted mutant frequencies using a pseudocount $\alpha$, such that $x_i^a(t) \rightarrow (1-\alpha) x_i^a(t) + \alpha/q$ for $i\in\{1,\ldots,L\}$ and $a\in\{1,\ldots,q\}$.  We chose $\alpha=0.2$ for all individual ensembles and both linear and B\'ezier methods.
}
\label{fig:comparison_linear_Bezier_HIV1}
\end{figure}

We applied multivariate B\'ezier interpolation to study human immunodeficiency virus (HIV-1) evolution in a set of 13 individuals\cite{liu2012vertical} (see
\hyperref[{subsec:negative_eigen}]{Methods} for details). 
The distribution of selection coefficients inferred using B\'ezier interpolation is highly correlated with previous analysis using linear interpolation\cite{sohail2021mpl}, indicating broad consistency with past results (\textbf{Fig.~\ref{fig:comparison_linear_Bezier_HIV1}}).
However, as we observed in simulations, inference using B\'ezier interpolation tends to result in slightly larger selection coefficients.
% TRUE BUT MAYBE UNNECESSARY TO NOTE HERE. ALSO A BIT TRICKY BECAUSE FLU ACTUALLY HAS A MUTATION RATE THAT IS NOT THAT DIFFERENT FROM HIV, BUT SARS-CoV-2's IS MUCH LOWER. -- Unlike other viruses such as seasonal flu and SARS-CoV-2, the HIV mutates rapidly and escapes from their host immune system\cite{sanjuan2010viral}. 
%Predicting what mutations contribute to virus prosperity or fitness, i.e., selection coefficients are clinically important task and is involved in the development of effective HIV vaccines\cite{altfeld2006hitting, barton2016relative, chakraborty2017rational}. 

Consistent with past analyses\cite{sohail2021mpl}, we found that the largest inferred selection coefficients are overwhelmingly associated with potentially functional mutations.
Among the largest 1\% of selection coefficients inferred across these 13 individuals, 
around 40\% correspond to mutations that help the virus to escape from the host immune system.
This represents a more than 20-fold enrichment in immune escape mutations among the most highly selected mutations, compared to chance expectations.

In summary, B\'ezier interpolation applied to real data leads to the inference of selection coefficients that are stronger than, but broadly consistent with, those that are found using linear interpolation. 
Large inferred selection coefficients also have clear biological interpretations.
For HIV-1, many highly beneficial mutations correspond to ones that the virus uses to escape from the immune system.

\subsection*{Recovery of rapidly decaying correlations underlies improved accuracy}\phantomsection\label{subsec:recovery_of_decaying_correlation}
To understand why MPL with B\'ezier interpolation yields more accurate inferences, we studied errors between true and estimated parameters as a function of the time interval $\Delta t$ between samples.
For arbitrary matrices $\mathrm{M}$ we define an error function $\mathcal{E}(\Delta t) = \| \mathrm{M}(\Delta t) - \mathrm{M}(1) \| / \| \mathrm{M}(1) \| $, normalizing by the matrix norm $\| \mathrm{M}(1) \|$, which corresponds to perfect sampling for the WF model.
In the discussion below we apply the $L_2$ norm, $\|M\| = \sqrt{\left(\sum_{i,j} M_{ij}^2\right)}$, but other conventions could also be considered.

\begin{figure}%[tbhp]
\centering
\includegraphics[width=\linewidth]{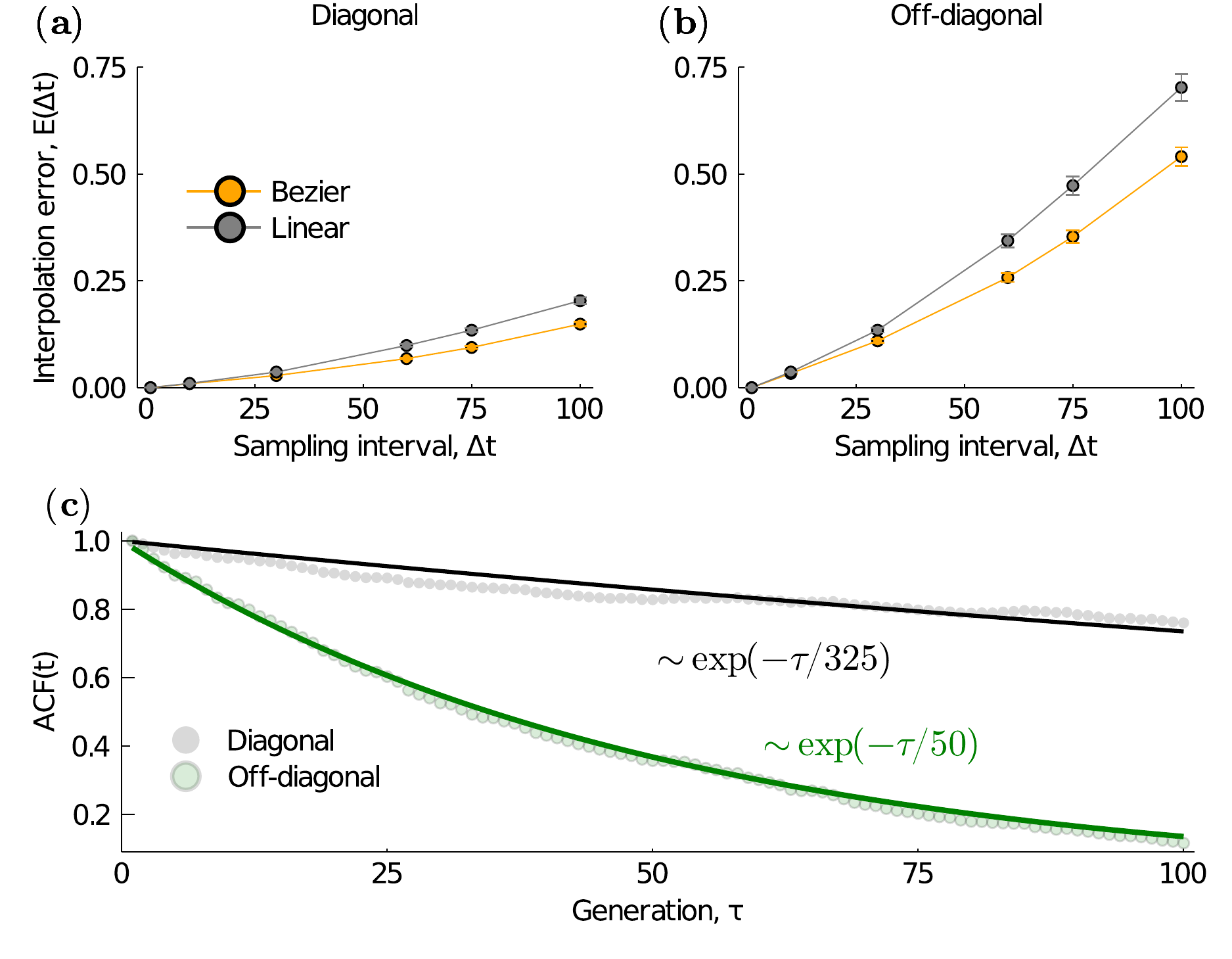}
\caption{ 
\textbf{The B\'ezier method suppresses interpolation error, especially for off-diagonal pairwise covariances.}
\textbf{(a)} Sampling time interval dependence for interpolation errors $\mathcal{E}(\Delta t)$ for diagonal covariances. \textbf{(b)} The same type of plot, but for off-diagonal pairwise covariances. We simulated WF dynamics using the model from \hyperref[{sec:wf_bezier}]{the previous section} and generated data sets that evolved to the 300th generation for each trial. For example, when $\Delta t=100$, results only use data from $t=0,100, 200, \text{ and } 300$. 
\textbf{(c)} The autocorrelation of off-diagonal covariance elements decays faster than diagonal ones.
%We numerically compute autocorrelations for diagonal and off-diagonal elements of the covariance matrix $C_{ij}=x_{ij}-x_ix_j$.
%The values of the exponent are around $1/325$ and $1/50$ for diagonal and off-diagonal ACFs, respectively. 
To simplify the analysis, we evaluated the autocorrelation function from generation $t = 50$. The diagonal autocorrelation shows non-monotonic decay after long times due to mutant frequencies that approach the frequency boundaries (i.e., 0 and 1).
\label{fig:fig5_interpolation_error}
}\end{figure}

Using the metric defined above, we found that B\'ezier interpolation yields better estimates for both the diagonal and off-diagonal terms of the mutant frequency covariance matrix. 
However, the error for the off-diagonal covariances is larger and increases much more rapidly with increasing $\Delta t$ than the error for the diagonal variances (\textbf{Fig.~\ref{fig:fig5_interpolation_error}a-b}).
The reduction in error for B\'ezier interpolation is more substantial for off-diagonal terms compared to diagonal ones.
Consistent with this observation, B\'ezier interpolation yields smaller improvements in performance for a simple version of MPL in which the off-diagonal terms of the integrated covariance matrix
%, which involve pairwise mutant frequencies, 
are ignored (\hyperref[{sup:MPL_SL_comparison}]{Methods}; referred to as the single locus (SL) method in ref.~\cite{sohail2021mpl}).

To study the time scale $\tau$ on which nonlinear effects become important and B\'ezier interpolation is advantageous, we modeled the covariance elements using a simple Langevin equation,  $\dot{z}(t) = -\lambda z(t) + \xi(t)$. 
Here $z(t)$ represents an element of the covariance matrix, $\lambda>0$ is a damping coefficient, and $\xi(t)$ is a standard white noise with $\langle \xi(t)\rangle=0$ and $\langle \xi(t)\xi(t+\tau)\rangle = 2\delta(\tau)$. 
Following this approach, a linear approximation should describe the evolution of $z(t)$ accurately if $\lambda \Delta t \ll 1$.  
The nonlinear nature of the $z(t)$ should become significant for $\lambda \Delta t  \sim 1$, and at this point the linear approximation cannot capture the actual evolution of $z(t)$.
Therefore, $\lambda \Delta t$ acts as a parameter that indicates whether linear interpolation is sufficient or inadequate.

The damping coefficient $\lambda$ can be estimated by computing the autocorrelation function (ACF) of the covariance matrix elements, which can be matched to expectations from the Langevin equation, $\langle x(t) x(t+\tau) \rangle \propto \exp(-\lambda \tau)$. 
In our simulations, the exponents of the ACF for diagonal and off-diagonal terms are around $\lambda_d \sim 1/325$ and $\lambda_o \sim 1/50$, respectively (\textbf{Fig.~\ref{fig:fig5_interpolation_error}c}). 
When the time between sampling events is $\Delta t = 75$, where B\'ezier interpolation clearly has an advantage (\textbf{Fig.~\ref{fig:fig4_PPV_Bezier_Linear}}), for diagonal and off-diagonal covariances 
%we have $\lambda_d \Delta t \sim 0.125$ and $\lambda_o \Delta t \sim 0.5$, respectively. 
we have $\lambda_d \Delta t \sim 0.23$ and $\lambda_o \Delta t = 1.5$, respectively. 
At this point, $\lambda_o \Delta t$ is $\mathcal{O}(1)$, indicating the onset of nonlinearity for off-diagonal terms. 
Consistent with this observation, for this value of $\Delta t$, B\'ezier interpolation has notably lower error for off-diagonal covariances than linear interpolation, while errors for the diagonal terms are comparable. 

While we focused specifically on the WF model in this example, the principle of autocorrelations and transitioning between linear and nonlinear behavior is general.
This can allow us to anticipate the benefit of nonlinear interpolation for a wide range of problems.

\subsection*{Inference of forces in Ornstein-Uhlenbeck processes}\phantomsection\label{sec:inverse_Ornstein_Uhlenbeck}
%To explore the potential benefits of B\'ezier interpolation for inference in other contexts, we applied this method to infer restoring forces in Ornstein-Uhlenbeck (OU) processes.
We further applied B\'ezier interpolation to accurately infer the collective forces in Ornstein-Uhlenbeck (OU) processes. 
Due to the mathematical simplicity and versatility of the OU process, it has played important roles in various fields such as physics, biology, and mathematical finance\cite{phillips2009maximum, bouchaud1998langevin, vasicek1977equilibrium, mamon2004three}. 
Data has been used to infer the parameters of OU processes describing phenomena including cell migration\cite{bruckner2019stochastic}, coevolution of species\cite{ho2014intrinsic}, and currency exchange rates \cite{roberts1998optimal}, to name a few examples.

\begin{figure}%[tbhp]
\centering
\includegraphics[width=\linewidth]{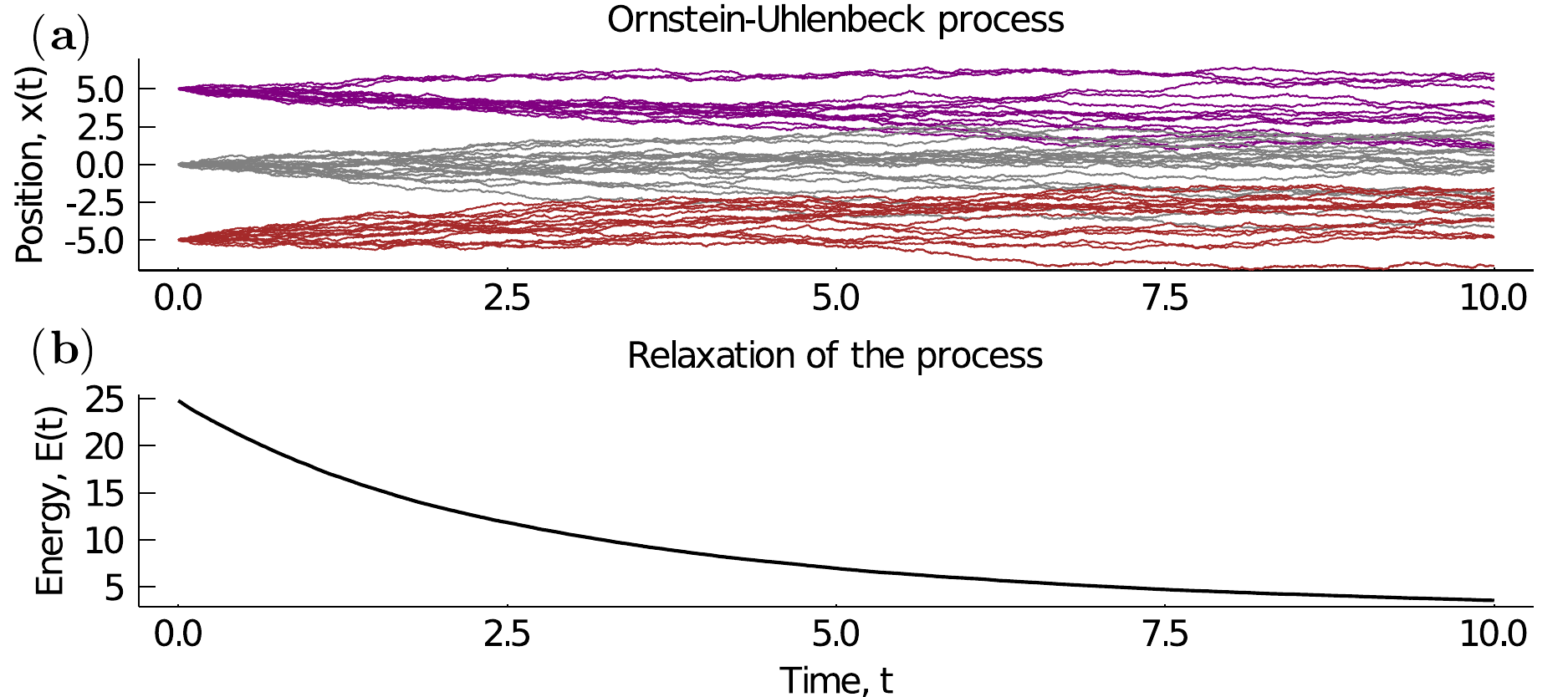}
\caption{ 
 \textbf{Typical Ornstein-Uhlenbeck dynamics.}
\textbf{(a)} We generated trajectories using the Euler-Maruyama (EM) scheme 5000 times with a small increment time step of $dt = 10^{-3}$. Each trajectory shows evolution of one of the elements of a multivariate variable. \textbf{(b)}  Evolution of the average effective energy of the OU process. The process relaxed from initial states randomly chosen from $\{-5,0,5\}^L$ to low energy states. We obtained the energy function, $E(t)=-\boldsymbol{x}(t)^\top \mathrm{J} \boldsymbol{x}(t) / 2$ by running EM simulations 100 times and averaging the results.
\label{fig:Energy_OU_process}}
\end{figure}

We consider the following OU process, a stochastic relaxation process of multivariate variables, 
\begin{equation}\label{eq:Ornstein_Uhlenbeck_main}
 d\boldsymbol{X}_t = \mathrm{J} \boldsymbol{X}_t + {\mathrm{\Sigma}}^{1/2} d\boldsymbol{W}_t \,. 
\end{equation}
Here $t$ is the time variable, $L$ is the number of OU stochastic variables, $\boldsymbol{X}_t \in \mathbb{R}^L $, $\mathrm{J} \in \mathbb{R}^{L\times L}$ is a negative semidefinite matrix, $\mathrm{\Sigma}$ is a time-independent noise covariance, and $\boldsymbol{W}_t$ is a Wiener process. 
We assume that the noise covariance matrix is constant over the evolution and given. 
Therefore, the unknown variable in the SDE in \eqref{eq:Ornstein_Uhlenbeck_main} is only the drift term, the interaction matrix $\mathrm{J}$.

%A certain stochastic process can be realized by a simple stochastic process, such as a Brownian process with a weight parameter. A similar idea is widely employed in computational physics and is known as importance sampling\cite{krauth2006statistical}. This weight variable that gives the conversion rule of the probability measure is called the \emph{likelihood ratio} \cite{risken1989fokker, liptser1977statistics}. In fact, the drift of the OU process is inferred by maximizing the Likelihood ratio for the entire stochastic process.
%\textcolor{red}{[The description in this next part is too sparse to follow; what likelihood ratio? what is a Radon-Nikodym derivative and what is it applied to? How does this solve the problem? Rather than mentioning the \emph{tools} that are used, I would try to describe the \emph{approach} to solving the problem.]} \textcolor{blue}{I introduced some sentences. } \emph{likelihood ratio} or \emph{Radon-Nikodym derivative} maximization facilitated by the Girsanov theorem\cite{karatzas2012brownian, liptser1977statistics}. 
One of the most commonly used approaches for inferring stochastic force in OU processes is maximizing the likelihood ratio or Radon-Nikodym derivative, which is the ratio of two probability measures\cite{risken1989fokker, liptser1977statistics}. Because of its ease of calculation and its mathematical rigor, this method is commonly employed in broad fields, such as mathematical finance\cite{phillips2009maximum}. In our problem, the likelihood ratio is defined as the probability density obeying the dynamics of \eqref{eq:Ornstein_Uhlenbeck_main} with interactions divided by the probability density of a ``null'' model with no interactions.
%Interestingly, we found that the maximum path likelihood method leads to exactly the same solution as the maximum likelihood ratio
Here, we inferred OU interactions by directly maximizing the path likelihood, as described for the WF model. Interestingly, this leads to exactly the same solution as the one for the standard likelihood/Radon-Nikdym derivative methods (\hyperref[sec:methods]{Methods}).

\begin{figure}%[tbhp]
\centering
\includegraphics[width=\linewidth]{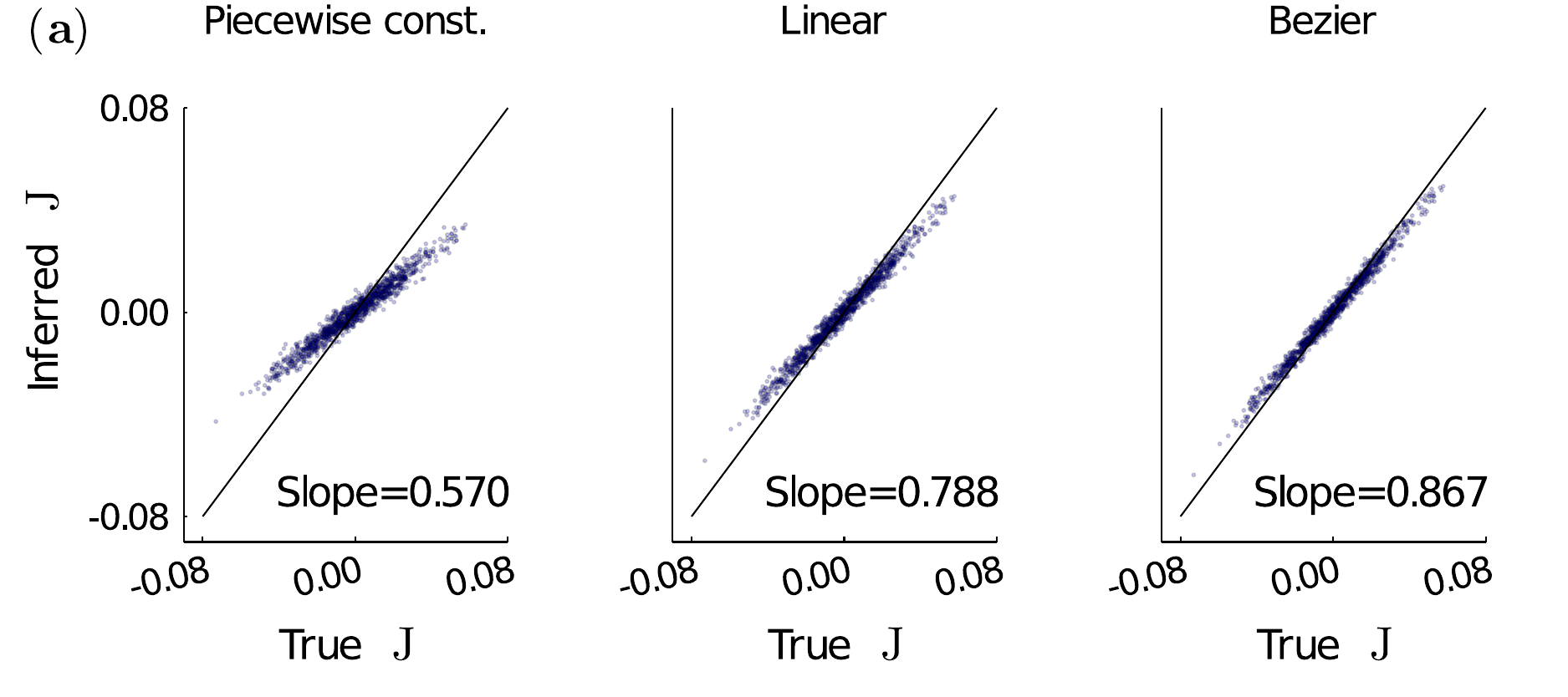}
\includegraphics[width=\linewidth]{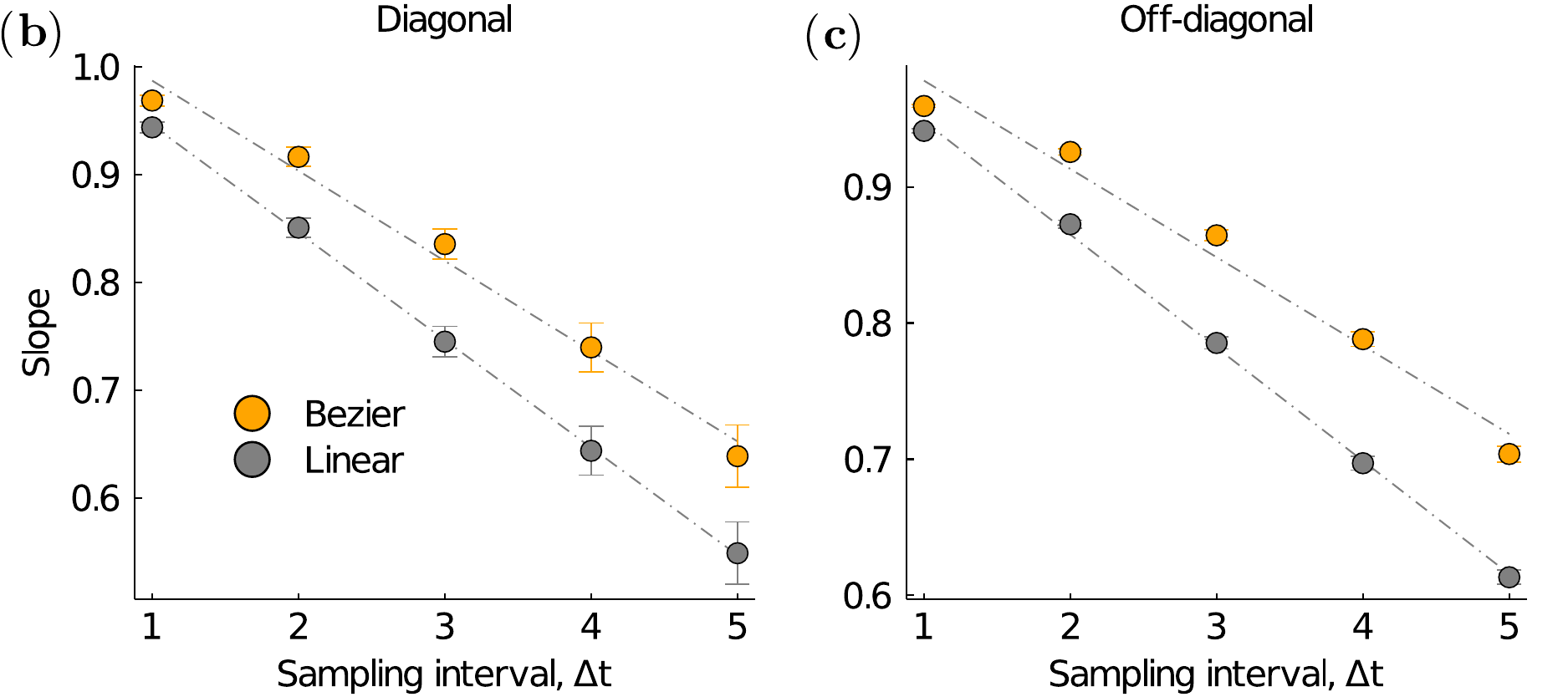}
\caption{ \textbf{Both linear interpolation and B\'ezier interpolation can improve the prediction accuracy of parameter predictions in the OU process.}
%\textcolor{red}{We can just use (a) for the top row and (b) for the bottom plot, here. The labels on the figure make it clear what is what. Also a small point, but the same thing is plotted on the horizontal and vertical axes of the (a) plots on the same scale, so the individual plots should be square rather than rectangular.} \textcolor{blue}{Ok, these are done.}  \textcolor{red}{ Let's also combine these into a single PDF instead of including separate PDFs, since this is how they will ultimately need to be sent to a journal.} \textcolor{blue}{These plots are actually in a single PDF. Do you suggest it to other figures?} \textcolor{blue}{Also, don't you think it is better to combine Fig.7 and Fig.8? These are associated with each other.}
\textbf{(a)} Comparison between true and inferred OU parameters using piecewise constant interpolation, linear interpolation, and B\'ezier interpolation. Regression slope values are included in each panel. Estimated interaction parameters using B\'ezier interpolation correspond most closely with the true parameters. 
%\textbf{(b)}  Parameters estimated with piecewise constant interpolation tend to underestimate particularly large values, while predictions using linear and B\'ezier interpolation do not tend to shrink. The prediction with the piecewise constant interpolation underestimates the parameters. The distributions of the true parameters and prediction with the B\'ezier interpolations are highly overlapped. 
\textbf{(b)} 
Dependence of the slope between true and inferred parameters on the time sampling interval $\Delta t = 1$, shown separately for the \textbf{(b)} diagonal and \textbf{(c)} off-diagonal interaction parameters of the $\mathrm{J}$ matrix.
In both cases, the slope decreases more gradually with increasing $\Delta t$ for B\'ezier interpolation than for linear interpolation.
\label{fig:scatter_estimated_couplings}}
\end{figure}

The interaction matrix $\hat{\mathrm{J}}$ that best describes the data is given by
\begin{align}\begin{aligned}
  \hat{\mathrm{J}} &= \left(\sum_{k=0}^{K-1} \Delta \boldsymbol{x}(t_k)\boldsymbol{x}(t_k)^\top \right) \\
  &\qquad\qquad \times\left(\sum_{k=0}^{K-1} \Delta t_k \boldsymbol{x}(t_k) \boldsymbol{x}(t_k)^\top \right)^{-1}\,. 
  \label{eq:MPLE_OU}
\end{aligned}\end{align}
%%%%%% Comment: the inverse of the sum of outer products acts from the right. We can check this by plugging the definition of the OU process to the analytical solution of the interaction matrix.
Here $\left( \boldsymbol{x}(t_k)\right)_{k=0}^{K-1}$ is the observed trajectory following the OU process, $\Delta t_k = t_{k+1} -t_{k}$ is an observation interval (not necessarily the same for all $k$), and $\Delta \boldsymbol{x}(t_k) = \boldsymbol{x}(t_{k+1}) - \boldsymbol{x}(t_{k})$ is the amount of change during the $k$th observation interval.

To generate test data, we simulated the OU process using negative definite interaction matrices parameterized as $\mathrm{J} = -\frac{\alpha}{\sqrt{P}}\sum_{\nu=1}^P \boldsymbol{\xi}_\nu \boldsymbol{\xi}_\nu^\top$.
This follows the construction of a Hopfield network, where $\boldsymbol{\xi}_\nu$ is a pattern generated from the multivariate normal distribution, $\boldsymbol{\xi}_\nu \sim \mathcal{N}(0,1)^L$, $\alpha = \mathcal{O}(1/L)$ is a small parameter, and $P$ is the number of embedded patterns. 
Hopfield networks were first constructed to study associative memory\cite{hopfield1982neural}, and have since been applied to problems such as the prediction of protein structure\cite{cocco2013principal, tubiana2019learning, shimagaki2019selection, shimagaki2019collective}.
This construction ensures that the OU process does not diverge.
We used the Euler-Maruyama (EM) scheme\cite{gillespie1996exact} to simulate \eqref{eq:Ornstein_Uhlenbeck_main} (\textbf{Fig.~\ref{fig:Energy_OU_process}a}).
We simulated 1000 trajectories each for 10 randomly generated interaction matrices, as described above. We chose $L=50$, and $\alpha =1/L=0.02$ in our simulations.
For inference, we sampled data from the simulations every $\Delta t = 1.0$ units of time.

Interaction parameters estimated using B\'ezier interpolation matched better with the true, underlying parameters than those inferred using linear interpolation or a piecewise-constant assumption for the $\boldsymbol{x}(t)$ (\textbf{Fig.~\ref{fig:scatter_estimated_couplings}a}).
In particular, large parameters inferred with linear interpolation or the piecewise-constant assumption tended to be underestimated.
In addition, we found that the slope relating the true and inferred parameters decreases as the sampling interval $\Delta t$ increases.
However, the slope between the inferred and true parameters decreases more slowly for B\'ezier interpolation compared to linear interpolation (\textbf{Fig.~\ref{fig:scatter_estimated_couplings}b-c}). 
Overall, OU interaction parameters inferred using B\'ezier interpolation more closely match the true, underlying parameters, than those inferred with simpler interpolation approaches or assumptions, with gains in performance that increase as data becomes more limited.

% We inferred parameters based on \eqref{eq:MPLE_OU}, using the numerator and denominator averaged over 1000 processes. To consider the effect of observation interval $\Delta t$, we masked data sets so that we can observe trajectories only for every $\Delta t$ step.
% The estimated interaction parameters using the maximum likelihood ratio with the piecewise constant, linear, and B\'ezier interpolations correlated with the true interaction parameters. However, the regression slope for the predicted parameters with the piecewise constant interpolation compared with the true parameter is smaller than the case of the parameters estimated with linear and B\'ezier interpolation (\textbf{Fig.~\ref{fig:scatter_estimated_couplings}a}), thus prediction with piecewise constant underestimates the parameters (\textbf{Fig.~\ref{fig:scatter_estimated_couplings}b}). 

% To more quantitatively investigate interpolation efficiency, we tested the dependency of the regression slope as a function of the observation interval. 
% We found that although both regression slopes of the predictions with the linear and B\'ezier interpolations decrease as increasing $\Delta t$, the slope of the prediction with the B\'ezier interpolation decreases more gradually than the slope of the prediction with the linear interpolation \textbf{Fig.~\ref{fig:accuracy_P_OU_model}}. Therefore, as the observation interval increase, the B\'ezier interpolation has more advantages than the linear interpolation.

\section*{Discussion}\phantomsection\label{sec:discussion}
Here we developed a nonlinear interpolation method based using B\'ezier curves that improves the inference of dynamical models from finite data. We applied our approach to two problems: the inference of natural selection in evolving populations and interactions in multivariate Ornstein-Uhlenbeck processes. B\'ezier interpolation makes inference more precise and reduces bias, especially for data sets that are more sparsely sampled. 

B\'ezier interpolation also has the advantage that it conserves sums of categorical variables, which is not typically guaranteed for standard stochastic regression methods such as Gaussian process regression/Kriging\cite{christensen2019advanced, mackay2003information} or nonlinear approaches such as kernel regression or least squares\cite{bishop2006pattern, mackay2003information}.
This property is especially useful for interpolating quantities that can be interpreted as probabilities (e.g., frequency vectors, as we considered above) or other conserved parameters. A few studies have applied regression methods to probabilities using logarithmic transformations. However, in such cases, regions around the 0 and 1 boundaries in the probability space tend to dominate regression results due to the coordinate transformation\cite{lin2020kriging}. %Therefore it is unsuitable for genetic selection prediction because it may not capture the bulk region where carries important information to tell beneficial/neutral/deleterious selection. -- sounds very reasonable but I'm not sure about stating this unless we know for sure!

Because of its generality, B\'ezier interpolation could be broadly applied to give more reliable results for dynamic inference problems. For example, our approach could be combined with methods to learn forces from non-equilibrium dynamics\cite{frishman2020learning, bruckner2020inferring}, or ones used to learn parameters of stochastic differential equations from finitely-sampled data\cite{iacus2008simulation, phillips2009maximum, ferretti2020building}.

%\begin{figure}%[tbhp]
%\centering
%\includegraphics[width=\linewidth]{figures/Slope_Dt-dependency.png}
%\caption{ 
%\textbf{The predictive regression slope using the B\'ezier interpolation %decreases more slowly than the gradient for linear interpolation as the %sampling time interval increases.}
%\textbf{(a)} Sampling interval dependency on the correlation between %predicted parameters and the true parameters in terms of the regression %slope. Samples are observed every $\Delta t = 1$ unit of time.
%  \textbf{(b)} The same type of plot for the off-diagonal interaction %parameters. 
%\label{fig:accuracy_P_OU_model}}
%\end{figure}

\begin{acknowledgements}
The work of K.S. and J.P.B.~reported in this publication was supported by the National Institute of General Medical Sciences of the National Institutes of Health under Award Number R35GM138233.
\end{acknowledgements}

\begin{contributions}
All authors contributed to methods development, data analysis, interpretation of results, and writing the paper. K.S.~performed simulations and computational analyses. J.P.B.~supervised the project.
\end{contributions}

\section*{References}
\bibliographystyle{naturemag}
\bibliography{Bezier_Interpolation}

\makeatletter
\renewcommand{\fnum@figure}{Supplementary Fig.~\thefigure}
\makeatother
\setcounter{figure}{0}

\section*{Methods}\phantomsection\label{sec:methods}
\subsection*{Data and code}

Raw data and code used in our analysis is available in the GitHub repository located at \url{https://github.com/bartonlab/paper-Bezier-interpolation}. This repository also contains Jupyter notebooks that can be run to reproduce the results presented here.

%Sometimes we write long equations that need to span both columns, like this
%\begin{widetext}
%\begin{align} \begin{aligned} \label{eq:action}
%P\left(\left(\bb{x}(t)\right)^T_{t=1} |\bb{x}(0)\right) &\approx %\left[\,\prod_{t=0}^{T-1} \frac{1}{\sqrt{\det C(\bb{x}(t))}}\left(\frac{N}{2\pi}\right)^{L/2}\,\prod_{i=1}^L d x_i(t+1) \right] \exp\left(-\frac{N}{2}S\left(\left(\bb{x}(t)\right)^T_{t=0}\right)\right),\\
%S\left(\left(\bb{x}(t)\right)^T_{t=0} \right) &= %\sum_{t=0}^{T-1}\sum_{i=1}^{L}\sum_{j=1}^{L}
%\left[x_i(t+1) - x_i(t) - d_i(\bb{x}(t))\right] %\left(C^{-1}(\bb{x}(t))\right)_{ij} \left[x_j(t+1) - x_j(t) - %d_j(\bb{x}(t))\right].
%\end{aligned} \end{align} 
%\end{widetext}

\subsection*{Optimization of control points for B{\'e}zier curves}
%We define the B{\'e}zier interpolation. 
%For simplicity, we will discuss a one-dimensional case, but the following %discussion can easily be extent to arbitrary dimensions.
%Suppose the number of data points we want to interpolate is $K+1$, and an interpolated global curve is drawn by smoothly connecting the fragments of the interpolation at each junction of the intervals.
%The B{\'e}zier interpolation can be formally defined as follows,
%\begin{equation}
%\begin{split}
%        B^{(k,k+1)}(\tau) &= \sum_{n=0}^3 \beta_n(\tau) \phi_{n}^{(k)}(\{x(t_{k'})\}) ~, \\
%        %\forall k &\in \{0,1,\ldots,K\} ~, \label{eq:Bezier_def}
%\end{split}
%\end{equation}
%where $\beta_n(\tau) = {3 \choose n}\tau^n(1-\tau)^{3-n}$ are the cubic Bernstein polynomial bases while $\tau\in [0,1]$, and $\phi_{n}^{(k)}(\{x(t_{k'})\})$ are the points that reflect the data dependency.
%To ensure that the segments at each interval $\{[t_k,t_{k+1}]\}_{k=1}^{K}$ are seamlessly connected, we impose the following conditions such that
%\begin{equation}
%    \phi_{0}^{(k)}(\{x(t_{k'})\}) =x(t_k),~~~\phi_{3}^{(k)}(\{x(t_{k'})\}) =x(t_{k+1}) ~.
%\end{equation}
%The other internal points, $\{ (  \phi_{1}^{(k)}, \phi_{2}^{(k)} ) \} $, are called control points and have the effect of controlling the outline of the interpolated curve. For the sake of simplicity, we will omit the arguments in the discussion as follows, unless there is no confusion. \\

For simplicity, we will discuss a one-dimensional case, but the following discussion can easily be extended to arbitrary dimensions.
The control points of B\'ezier curves are obtained by solving an optimization problem that is derived from properties we want the B{\'e}zier curve to satisfy. In this study, we impose the $C^2$ smoothness condition, which is that up to the second derivative of the curve exist. Formally, we can represent these conditions as follows,
\begin{equation}
    \partial_\tau x_B^{(k-1)}(\tau=1) = \partial_\tau x_B^{(k)}(\tau=0) ~, \label{eq:Bezier_constraints1}
\end{equation}
and, 
\begin{equation}
    \partial^2_\tau x_B^{(k-1)}(\tau=1) = \partial^2_\tau x_B^{(k)}(\tau=0) ~, \label{eq:Bezier_constraints2}
\end{equation}
Where, $x_B^{(k)}(\tau)$ is the interpolated function between successive discrete time points $t_k$ and $t_{k+1}$ and defined in \eqref{eq:Bezier_interpolation_simple}.
Since these constraints are defined at each junction of adjacent segments, the number of conditions is $2(K-1)$. On the other hand, the number of control points is $2K$, so we will introduce two more constraints to make the problem solvable:
\begin{equation}
    \begin{split}
        \partial^2_\tau x_B^{(0)}(\tau=0) &= 0 \\
        \partial^2_\tau x_B^{(K-1)}(\tau=1) &= 0 ~ \nonumber \label{eq:Bezier_boundary_condition}
    \end{split}~. 
\end{equation}

By rearranging \eqref{eq:Bezier_constraints1} and  \eqref{eq:Bezier_constraints2}, we can reduce them to the following difference equations.

\begin{equation}
         \phi_1^{(k)} - 2 x^{(k)} = \phi_2^{(k-1)} ~,  \label{eq:Bezier_constraints1_simple}
 \end{equation}
 and
 \begin{equation}
     -2 \phi_1^{(k)} + \phi_2^{(k)}  = \phi_1^{(k-1)} - 2 \phi_2^{(k-1)} ~.  \nonumber \label{eq:Bezier_constraints2_simple}
 \end{equation}
 
Also, the additional boundary constraints lead to
 \begin{equation}
    \begin{split}
        x^{(0)} -2 \phi_1^{(0)} + \phi_2^{(0)} &= 0 ~, \\
        \phi_1^{(k-1)} - 2 \phi_2^{(k-1)} + x^{(k)} &= 0  \nonumber \label{eq:Bezier_boundary_condition_simple}
    \end{split}~. 
 \end{equation}
 
These difference equations are summarized as the following single linear equation by assuming that $\{\phi_2^{(k)}\}_{k=0}^{K-1}$ is a function of $\{\phi_1^{(k)}, x^{(k)}\}_{k=0}^{K-1}$, then by marginalizing $\{\phi_2^{(k)}\}_{k=0}^{K-1}$ from the difference equations,
\begin{equation}
\mathrm{M}^{\rm{Bez}, K} \boldsymbol{\phi}_1  = \boldsymbol{\psi}((x^{(k)})_{k=0}^{K+1}) \label{eq:linear_eq_Bez} ~,
\end{equation}
where $\boldsymbol{\phi}_1 = (\phi_1^{(0)},\ldots,\phi_1^{(K)})^T$, and let 
\begin{equation}
\boldsymbol{\psi}((x^{(k)})_{k=0}^{K+1}) =
\begin{pmatrix}
    x^{(0)}+2x^{(1)} \\
    2(2x^{(1)}+x^{(2)})  \\
    \vdots \\
    2(2x^{(K-1)}+x^{(K)})  \\
    8x^{(K)} + x^{(K+1)} \\
\end{pmatrix} ~, \label{eq:Bezier_linear_eq_rhs}
\end{equation} 
and the matrix $\mathrm{M}_B^{(K)} $ is defined as
\begin{equation}
        \mathrm{M}_B^{(K)}    = 
    \begin{pmatrix}
2     &      1 &       0 &  \dots  &  \dots  & \dots   &  \dots & 0      \\
1     &      4 &       1 &       0 &  \dots  &  \dots  &  \dots & \vdots \\
0     &      1 &       4 &       1 &     0   &  \dots  & \ldots & \vdots \\
\vdots&  \dots & \ddots  & \ddots  &  \ddots &  \dots  & \ldots & \vdots \\
\vdots&  \dots &  \dots  & \ddots  &  \ddots &  \ddots & \ldots & \vdots \\
\vdots&  \dots &  \dots  &      0  &       1 &     4   &      1 & 0      \\
\vdots&  \dots &  \dots  &  \dots  &       0 &     1   &      4 & 1      \\
0&  \dots &  \dots  &  \dots  &  \dots  &     0   &      2 & 7      \\
\end{pmatrix}   ~. \label{eq:matrix_Bezier}
\end{equation}
By solving \eqref{eq:linear_eq_Bez}, we get a set of control points, hence we get a B{\'e}zier curve.
Interestingly, instead of the $C^2$ smoothness constraint, assuming a $C^1$ smoothness condition and imposing a constraint that minimizes the Euclidean distance of the total trajectory leads to almost the same linear equation in \eqref{eq:linear_eq_Bez} depending on \eqref{eq:matrix_Bezier} and \eqref{eq:Bezier_linear_eq_rhs}.\\

For multivariate frequencies, the B{\'e}zier curve can be obtained by solving each linear equation individually. Practically, the control points are obtained by operating the inverse of $\mathrm{M}_B^{(K)}$ to $\boldsymbol{\psi}((x^{(k)})_{k=0}^{K+1})$ vectors on each site $i\in\{1,\ldots,L \}$. Thus, we can efficiently perform the operation and its computational time is fast.
Also, the above arguments are held for the arbitrary $q>1$ dimension case, which is relevant, for example, when considering the frequency of individuals with multiple possible nucleotides or amino acids at each site in a genetic sequence. Replacing scalar variables with vector variables leads to exactly the same linear equation in \eqref{eq:linear_eq_Bez}.

\subsection*{Integrated frequency and covariance using B{\'e}zier interpolation}
In this section, we will show explicit representations of the integrated mutant frequencies and covariances from the WF model using B{\'e}zier interpolation.

To derive it, we apply the following useful properties of $P$-th order Bernstein basis ($P=3$ for quadratic B{\'e}zier interpolation), for $\forall n\in \{0,1,\ldots,P\}$, 
\begin{equation}
    m_n^{(P)} := \int_0^1 \beta_n(\tau) {\rm d}\tau= \frac{1}{P+1} ~, \label{eq:property_Bernstein1}
\end{equation}
and, for $\forall n,m\in \{0,1,\ldots,P\}$,
\begin{equation}
    \mathrm{Q}_{n m}^{(P)} := \int_0^1 \beta_n(\tau) \beta_m(\tau) {\rm d}\tau
    = \frac{ 
    {P \choose n} {P \choose m}
    }
    {
    (2P+1){2P \choose n+m}
    } ~. \label{eq:property_Bernstein2}
\end{equation}
More general properties of the Bernstein basis can be found in refs.~\cite{doha2011integrals, alturk2016application}.

First, we will get the integrated single mutant frequency at site $i$, which is shown below,
\begin{equation}
\begin{split}
    \Delta x_{B,i}^{(k)} &:= \Delta t_k \int_0^1 x_{B,i}^{(k)} (\tau) {\rm d}\tau \\
    &= ~ \Delta t_k \sum_{n=0}^P \left( \int_{0}^1   \beta_n(\tau) {\rm d}\tau \right) \phi_{i,n}^{(k)} \\
    &=  ~ \Delta t_k \sum_{n=0}^P m_n^{(P)} \phi_{i,n}^{(k)} \\
    &= \frac{1}{(P+1)}\sum_{n=0}^P \phi_{i,n}^{(k)} ~, \label{eq:integrated_Bezier_single} 
\end{split}
\end{equation}
we used the property of Bernstein in \eqref{eq:property_Bernstein1}.

Next, we will get the integrated covariance for different sites at $i$ and $j$,
\begin{equation}
    %\begin{split}
        \Delta C^{(k)}_{ij} :=
    \Delta t_k \int_0^1 
    \left(
        x_{B,ij}^{(k)}(\tau) - ~ x_{B,i}^{(k)}(\tau) x_{B,j}^{(k)}(\tau) \right) {\rm d}\tau ~,\label{eq:integrated_Bezier_double} 
    %\end{split}
\end{equation}
the first term in \eqref{eq:integrated_Bezier_double} is the same as in \eqref{eq:integrated_Bezier_single} but we replaced a single interpolated mutant frequency by a matrix that contains the entire interpolated pairwise mutant frequencies as its elements. 

The second term of the covariance in \eqref{eq:integrated_Bezier_double} is also straightforward,
\begin{equation}
    \begin{split}
        &\int_{0}^1 x_{B,i}^{(k)}(\tau) x_{B,j}^{(k)}(\tau) {\rm d}\tau \\
        &= \int_{0}^1  
        \left( \sum_{n=0}^P\beta_n(\tau)\phi_{i,n}^{(k)}  \right) 
        \left( \sum_{m=0}^P\beta_m(\tau)\phi_{j,m}^{(k)}  \right) {\rm d}\tau \\
        &= \sum_{n=0}^P\sum_{m=0}^P \left( \int_{0}^1 \beta_n(\tau)\beta_m(\tau) {\rm d}\tau \right) \phi_{i,n}^{(k)} \phi_{j,m}^{(k)} \\
        &= \sum_{n=0}^P\sum_{m=0}^P Q_{n m}^{(P)} \phi_{i,n}^{(k)} \phi_{i,m}^{(k)} ~.  \nonumber \label{eq:Bez_cov_2nd}
    \end{split}
\end{equation}
Here we used the property of Bernstein \eqref{eq:property_Bernstein2} in the last equality.

In the case of the $P=3$, which is the cubic B{\'e}zier, 
$\mathrm{Q}^{(3)}$ matrix will be
\begin{equation}
    \mathrm{Q}^{(3)} = \begin{pmatrix}
            \alpha & \beta & \gamma & \delta \\
            \beta & \gamma & \delta & \gamma\\
            \gamma & \delta & \gamma & \beta \\
            \delta & \gamma & \beta & \alpha
        \end{pmatrix} ~,  \nonumber
\end{equation}
where $\alpha=1/7, \beta=1/14, \gamma=1/35, \delta=1/140$.\\

\subsection*{Normalization of probabilities}\label{subsec:normalization}
%In this section, we will discuss that the integrated covariance can actually satisfy the properties of a covariance matrix when the B{\'e}zier interpolation is used for the interpolation of the probability distribution.
% this section, we will discuss characteristic properties of the B{\'e}zier interpolation that the interpolated trajectories of the frequency distributions satisfies properties as a probability distribution.
%More specifically, we will address the following properties:  the \emph{normalizability} that ensures that a sum of the interpolated function of all states is always 1, and \emph{positive semidefiniteness} of the integrated covariance. \\
We will show that the interpolation of probability trajectories using the B{\'e}zier interpolation is always normalized. We refer to this property as normalizability, hereafter.

First, we will discuss the normalizability of the interpolated probability distribution for a categorical distribution depending on an arbitrary number of states $q>1$.
Next, we denote a probability distribution depending on the data points $k$ and index $i$ as $\boldsymbol{x}^{(k)}_i = (x^{(k)}_{i,1}, \ldots, x^{(k)}_{i,q})^T$, and a sum of the all states is normalized, that is $\sum_{a=1}^q x_{i,a}^{(k)}=1$ for all $k,i$. 

Then, we can prove that when probability distributions are interpolated using B{\'e}zier's method, any interpolated function $\boldsymbol{x}_{B, i}{(k)}= (x_{B,i,1}^{(k)}, \ldots, (x_{B,i,q}^{(k)})^\top$ is also normalized in arbitrary point $\tau \in [0,1]$:
\begin{equation}
    \sum_{a=1}^q x_{B,i,a}^{(k)} = 1 ~.  \nonumber
\end{equation}\\

For the sake of simplicity, we will omit the site index hereafter.
To see the proof, we will start by showing the normalizability of the control points $ \sum_{a=1}^q \phi_{1,a}^{(k)}=1, \forall k$ ~ because this condition immediately leads to $\sum_{a=1}^q \phi_{2,a}^{(k)}=1$ by plugging it into the \eqref{eq:Bezier_constraints1_simple}, and the following part is straightforward as shown below, 
 \begin{equation}
 \begin{split}
     \phi_{2,a}^{(k-1)} &= 2x_{a}^{(k)} - \phi_{1,a}^{(k)} \\
     &= 2(1- \sum_{b=1 | \neq a}^q x_b^{(k)}) - ( 1 - \sum_{b=1|\neq a}^q \phi_{1,b}^{(k)}) \\
     &=1 - (2\sum_{b=1 | \neq a}^q x_b^{(k)} -  \sum_{b=1|\neq a}^q \phi_{1,b}^{(k)} ) \\
     &= 1- \sum_{b=1|\neq a}^q \phi_{2,b}^{(k-1)} ~,  \nonumber \label{eq:relation_a_b_1}
 \end{split} 
 \end{equation}
 so $\sum_{a=1}^q {\phi}_{2,a}^{(k)}=1$ and it is normalized when $\boldsymbol{\phi}_{1}^{(k)}$ is normalized for $k\in \{ 1,\ldots, K-1 \}$. 
 In the case of boundaries, time points at $k = 0, K$, exactly the same argument holds, which is almost trivial, so we omit to repeat the same kind of proof.
 
Therefore, we will show the normalizability of $\boldsymbol{\phi}_{1}^{(k)}$ as follows. 
First, we consider a sum of all the states on the left hand side in \eqref{eq:Bezier_linear_eq_rhs}, 
\begin{equation}
     l.h.s. = \mathrm{M}_B^{K}
        \begin{pmatrix}
    \sum_{a=1}^q\phi_{1,a}^{(0)} \\
    \vdots \\
    \sum_{a=1}^q\phi_{1,a}^{(K)} \\
    \end{pmatrix} ~.  \nonumber
\end{equation}
Next, we also perform a sum of all the states on the right hand side in \eqref{eq:Bezier_linear_eq_rhs},
\begin{equation}
        r.h.s. = \sum_{a=1}^q 
    \begin{pmatrix}
    x^{(0)}+2x^{(1)} \\
    2(2x^{(1)}+x^{(2)})  \\
    \vdots \\
    2(2x^{(K-1)}+x^{(K)})  \\
    8x^{(K)} + x^{(K+1)} \\
    \end{pmatrix} = 
    \begin{pmatrix}
    3 \\
    6  \\
    \vdots \\
    6  \\
    9 \\
    \end{pmatrix} ~. \nonumber
\end{equation}
Then, we immediately notice that 
\begin{equation}
    \mathrm{M}_B^{K} \boldsymbol{1} = (3,6,\ldots,6,9)^\top ~. \nonumber
    %\begin{pmatrix}
    %3 \\
    %6  \\
    %\vdots \\
    %6  \\
    %9 \\
    %\end{pmatrix} ~.
\end{equation}
%$\hat{\mathrm{M}}^{\rm{Bez}, K} \boldsymbol{1}$ is exactly the same as the result we obtained by summing the all states of the right side of \eqref{eq:Bezier_linear_eq_rhs}.
Therefore, we find the normalization of the control points $\sum_{a=1}^q\phi_{1,a}^{(k)} = 1, ~\forall k\in \{0,1,\ldots,K \} $.

Finally, we sum the interpolated function using B\'ezier's method at arbitrary $\tau$ while considering the normalizability conditions for the control points we have seen earlier.
%when the normalization conditions for $\boldsymbol{x}^{(k)}, \forall k$ is satisfied, then the normalization conditions for $\boldsymbol{\phi}_{1}^{(k)},\boldsymbol{\phi}_{2}^{(k)} \forall k$ are also satisfied, 
A sum of the interpolated functions for the all states $a\in \{1,\ldots,q \}$ at any position $\tau \in [0,1]$ is:
\begin{equation}
    \sum_{a=1}^q x_{B,i,a}^{(k)} = \sum_{l=0}^P \beta_l(\tau) = 1 ~,  \nonumber\label{eq:proof_end_Bezeir_norm}
\end{equation}
for the first equality, we used the fact that all the control points are normalized. 
For the second equality, we used the nature of the Bernstein polynomial, a sum of all the Bernstein bases is one.

\subsection*{Treatment for negative interpolated frequencies and negative eigenvalues in real data}\label{subsec:negative_eigen}
The sum of $q$ categorical variables using B\'ezier interpolation is conserved, guaranteeing the conservation of probability density. However, interpolated probabilities can occasionally exceed the boundaries at 0 and 1, and eigenvalues of the integrated covariance matrix can become negative. This issue can occur when frequency trajectories are close to the boundaries, variables take one of the multiple possible states ($q>0$), and sampling points are heterogeneously and sparsely distributed. 

To alleviate this problem, we employed the following treatment: if the time interval $\Delta t_k  = t_{k+1}-t_k$ is greater than a threshold value (set to 50 days for the analysis of HIV-1 sequence data), then we insert mean frequency points at the middle time points $(t_{k+1}+t_k)/2$ such that $\boldsymbol{x}(t_k) + \boldsymbol{x}(t_{k+1})/2$ . 
In addition, for each frequency individually, we insert mean frequency points at middle time points when the frequency changes sharply within one time interval (more than 70\% change in the case of HIV-1 data).
%Another approach, adjusting mutant frequencies using a pseudocount $\alpha$, such that $x_i^a(t) \rightarrow (1-\alpha) x_i^a(t) + \alpha/q$ for $i\in\{1,\ldots,L\}$ and $a\in\{1,\ldots,q\}$ can prevent the negative eigenvalue problem. However, we did not employ this method in our study because information about whether or not mutant frequencies are near the boundaries plays an important role in estimating selection coefficients (see \eqref{eq:sMAP} in the main text).

\subsection*{Maximum path-likelihood estimation for the Ornstein-Uhlenbeck process}
Based on the stochastic differential equation (STD) defined in \eqref{eq:Ornstein_Uhlenbeck_main}, We can get the following Fokker-Planck equation \cite{risken1989fokker}, which is characterized by the drift and diffusion terms, 
%\begin{widetext}
\begin{equation}
    \begin{split}
        &\frac{\partial}{\partial t} p(\boldsymbol{x}(t),t) 
        = \mathcal{L}~ p(\boldsymbol{x}(t),t) \\
        \mathcal{L} &= 
        - \sum_{i=1}^L\sum_{j=1}^i J_{ij}x_j  \frac{\partial}{\partial x_i} 
        +\sum_{i,j=1}^L  {\mathrm{\Sigma}}_{ij}  \frac{\partial^2}{\partial x_i \partial x_j} \,.  \label{eq:FP_eq_pairwise}
    \end{split}
\end{equation}
%\end{widetext}
The  first term corresponds to the drift due to the pairwise interaction, and the second term corresponds to the diffusion due to the white noise.

The FP equation in \eqref{eq:FP_eq_pairwise} is effectively a diffusion equation for probability measures, and the general solution of the diffusion equation characterized by the drift and diffusion terms is known and defined as a transition probability between time points $t_k$ and $t_{k+1}=t_k+\Delta t_k$,
\begin{align}\begin{aligned}
  &p\left(\boldsymbol{x}(t_{k+1}), t_{k+1} ~|~\boldsymbol{x}(t_k), t_k \right)\\
  & = 
\frac{1}{\sqrt{\| 2\pi\Sigma\| \Delta t_k}}
\exp\Big( 
 -\frac{1}{2\Delta t_k} 
    \left(
    \Delta \boldsymbol{x}(t_{k}) - \Delta t_k\mathrm{J} \boldsymbol{x}(t_k) 
    \right)^\top
    \\
&\qquad\qquad \times \Sigma_t^{-1}  \left(
    \Delta \boldsymbol{x}(t_{k}) - \Delta t_k \mathrm{J} \boldsymbol{x}(t_k) 
    \right)    
    \Big)\,,
\end{aligned}  \nonumber \end{align}
where $\Delta \boldsymbol{x}(t_k) = \boldsymbol{x}(t_{k+1}) - \boldsymbol{x}(t_k)$. 
The solution of the FP equation tells that as the time interval approaches zero, the transition probability goes to the \emph{Kronecker delta} like distribution having a finite probability density around the previous time step. As the time interval increase, the variance increase as the square root of time, which is the nature of Brownian diffusion.
 
The likelihood path function for the OU model can be defined as a product of the transition probability because of the independence of the increments of the Wiener processes. Hence the log path-likelihood can be written as
\begin{align}\begin{aligned}
  &\mathcal{S}(\mathrm{J} |\Gamma( ( \boldsymbol{x}(t_k) )_{k=0}^{K-1} )) \\
  &= \sum_{k=0}^{K-1} \left( 
 -\frac{1}{2\Delta t_k} 
    \left(
    \Delta \boldsymbol{x}(t_{k}) - \mathrm{J} \boldsymbol{x}(t_k) 
    \right)^\top
    \Sigma_t^{-1}
\left(
    \Delta \boldsymbol{x}(t_{k}) - \mathrm{J} \boldsymbol{x}(t_k) 
    \right)    
    \right)\\
    &\qquad  + {\rm const.} ~\label{eq:action_OU_model}
\end{aligned}\end{align}
The log-likelihood corresponds to the \emph{action} in statistical physics, 
where $\Gamma( ( \boldsymbol{x}(t_k) )_{k=0}^{K-1} ) = (\boldsymbol{x}(t_0), \ldots, \boldsymbol{x}(t_{K-1}))$ is a single trajectory of the stochastic variable. 

Since the action in \eqref{eq:action_OU_model} is a convex function of the coupling matrix, the most probable coupling matrix (i.e., the one that maximizes the likelihood of the observed path) can be obtained by computing the derivative of the action with respect to the coupling matrix, setting it to zero, and solving for the coupling matrix. 

%\begin{equation}
%    \frac{\partial}{\partial \hat{J}} \mathcal{S}(\hat{J} | \Gamma(\{ \boldsymbol{x}(t_k) \}_{k=0}^{K-1} )) = \sum_{k=0}^{K-1} \hat{\Sigma}^{-1} (\Delta \boldsymbol{x}(t_k) - \hat{J} \boldsymbol{x}(t_k) )\boldsymbol{x}(t_k)^\top \to \hat{0} ~, \label{eq:derivative_action_OU}
%\end{equation}
 
The derivative of the log-path-likelihood function with respect to the coupling matrix can be factorized by the noise covariance because of its time-independence, giving the following closed-form solution
%\begin{equation}
%    \begin{split}
%        \hat{\mathrm{J}} =  \left(\sum_{k=0}^{K-1} \Delta %\boldsymbol{x}(t_k)\boldsymbol{x}(t_k)^\top  \right)\left(\sum_{k=0}^{K-1} %\boldsymbol{x}(t_k) \boldsymbol{x}(t_k)^\top  \right)^{-1} ~. \label{eq:MPLE_OU}
%    \end{split} 
%\end{equation}
\begin{align}\begin{aligned}
  \hat{\mathrm{J}} &= \left(\sum_{k=0}^{K-1} \Delta \boldsymbol{x}(t_k)\boldsymbol{x}(t_k)^\top \right) \\
  &\qquad\qquad \times\left(\sum_{k=0}^{K-1} \Delta t_k \boldsymbol{x}(t_k) \boldsymbol{x}(t_k)^\top \right)^{-1}, 
  \label{eq:MPLE_OU_method}
\end{aligned}\end{align}

The single trajectory maximum path likelihood estimate (MPLE) in \eqref{eq:MPLE_OU_method} can be easily generalized to the case of multiple trajectories or paths by replacing the action in \eqref{eq:action_OU_model} to an ensemble-averaged action $\langle \mathcal{S}(\mathrm{J} |\Gamma)  \rangle_{\Gamma \in {\rm ensemble}} $ (or, equivalently, by observing that the likelihood of a set of independent paths is equal to the product of the likelihoods for each individual path). The corresponding MPLE solution after ensemble averaging is 
\begin{align}\begin{aligned}
  \hat{\mathrm{J}} &= \left(\sum_{m=1}^M\sum_{k=0}^{K^m-1} \Delta \boldsymbol{x}^m(t_k)\boldsymbol{x}^m(t_k)^\top  \right) \\
  &\qquad\qquad \times\left(\sum_{m=1}^M\sum_{k=0}^{K^m-1} \Delta t_k\boldsymbol{x}^m(t_k) \boldsymbol{x}^m(t_k)^\top  \right)^{-1}, 
  \label{eq:MPLE_OU_ensemble}
\end{aligned}  \nonumber\end{align}
where $m=1,\ldots, M$ is the ensemble index.\\

In fact, by assuming the discretization of \eqref{eq:Ornstein_Uhlenbeck_main}, we can estimate sample size dependence on the MPLE, and it is an \emph{unbiased estimator}, as shown in below,
\begin{align}\begin{aligned}
  \hat{\mathrm{J}} &= \left(\sum_{m=1}^M\sum_{k=0}^{K^m-1} \Delta t_k\left( 
        \hat{\mathrm{J}}^{*} \boldsymbol{x}^m(t_k) + \boldsymbol{W}(t_k)
        \right)\boldsymbol{x}^m(t_k)^\top  \right) \\
  &\qquad\qquad \times\left(\sum_{m=1}^M\sum_{k=0}^{K^m-1} \Delta t_k \boldsymbol{x}^m(t_k) \boldsymbol{x}^m(t_k)^\top  \right)^{-1}\\
  &\sim \hat{\mathrm{J}}^{*} + \Tilde{\boldsymbol{W}} / \sqrt{M }  ~~~ %\underrightarrow{{}_{M\to \infty}} ~~~ \hat{\mathrm{J}}^{*} ~,
  \xrightarrow{M\to \infty}~ \hat{\mathrm{J}}^{*}\,.
\end{aligned}  \nonumber\end{align}
To derive the scaling of the estimation bias, we used the assumption of the independence of the white noise. \\

\subsection*{Cameron-Martin-Girsanov theorem and application for Ornstein-Uhlenbeck process inference}
In this section, we will show that the inference problem of the OU model can be solved by maximizing the \emph{Radon-Nikodym} (RN) derivative or \emph{likelihood ratio}, which is facilitated by the Cameron-Martin-Girsanov (CMG) theorem  \cite{cameron1944transformations,girsanov1960transforming, risken1989fokker, liptser1977statistics}. Since the aim of this section is only to rationalize the inference approach based on the CMG theory, we will discuss minimal ingredients of the CMG theory. A more general and comprehensive description can be found in refs.~\cite{risken1989fokker, liptser1977statistics}.\\

First, let us define the RN derivative. If two probability measures $\mathbb{P}$ and $\mathbb{Q}$ satisfy the following conditions, then the $\mathbb{P}$ and $\mathbb{Q}$ are said to be \emph{mutually absolutely continuous},
\begin{equation}
    \begin{split}
        \mathbb{E}_\mathbb{Q}[Y] &= \mathbb{E}_\mathbb{P}[YZ] \\
        \mathbb{E}_\mathbb{P}[Y] &= \mathbb{E}_\mathbb{Q}[Y/Z]~,
    \end{split}  \nonumber
\end{equation}
where $\forall Y>0$. $Z$ is some random variable, and if it satisfies the condition, $\mathbb{E}_\mathbb{P}[Z]=1$, then $Z$ is called Radon-Nikodym derivative (or likelihood ratio). In fact, it is nothing more than the changing of the probability measures 
\begin{equation}
    \mathbb{E}_\mathbb{Q}[Y] = \int Y d\mathbb{Q} = \int Y \frac{d\mathbb{Q}}{d\mathbb{P}} d\mathbb{P} =  \mathbb{E}_\mathbb{P}\left[Y\frac{d\mathbb{Q}}{d\mathbb{P}} \right]\,.  \nonumber
\end{equation}
Therefore, such a random variable $Z$ is denoted as $\frac{d\mathbb{Q}}{d\mathbb{P}} := Z$ in general. 
Since the RN derivative gives transformation of a probability measure to another probability measure without obtaining (or even knowing explicit form of) the probability measure $\mathbb{Q}$, it enables us to estimate some statistics under the probability measure $\mathbb{Q}$ that are unobtainable directly. For example, \emph{importance sampling} falls in this class of problems and is widely used in computational studies. \\

Informally speaking, the CMG theorem states that under some transformation of the drift term in a Wiener process, a probability measure after the transformation exists and can represent its explicit RN derivative. So, the CMG theorem provides a way to estimate the statistics under a probability density after a general transformation of the drift of the Wiener process.\\

More formally, the statement of the Cameron-Martin-Girsanov theorem is that 
for a Brownian motion $\{W_t \}_{t\geq 0}$ that follows a probability measure $\mathbb{P}$ and observable process $\gamma_t$ that satisfies the following \emph{Nikodym condition} 
\begin{equation}
    \mathbb{E}_{\mathbb{P}}\left[ \exp\left( 
    \frac{1}{2}\int_0^t \gamma_s^2 ds
    \right)\right] < \infty~, ~~ \forall t \geq 0 ~,  \nonumber \label{eq:Nikodym_condition} 
\end{equation} 
the probability measure $\mathbb{Q}$ that corresponds to the stochastic process 
$dX_t = -\gamma_t dt +dW_t$ 
\footnote{
We can transform most stochastic processes to this type of stochastic process. For example, a stochastic process given by 
\begin{equation}
    dX_t = \gamma_t dt + \sigma_t({X}_t) dW_t~, \nonumber
\end{equation}
Here, $\sigma_t({X}_t)$ is a covariance that can depend not only on time but also on random variables, so it becomes a multiplicative noise \cite{liptser1977statistics}.
Then we transform the stochastic process and drift such that $\tilde{X}_t = \sigma_t({X}_t)^{-1} X_t$ and $\tilde{\gamma}_t({X}_t) = \sigma_t({X}_t)^{-1} \gamma_t$, then we can get the following stochastic process 
\begin{equation}
    d\tilde{X}_t = \tilde{\gamma}_t dt + dW_t~. \nonumber
\end{equation}
} exists and the $\mathbb{Q}$-process is equivalent to $\mathbb{P}$-Brownian motion by modifying the Wiener process such that
\begin{equation}
    \Tilde{W}_t = W_t + \int_0^t \gamma_s ds~  \label{eq:Girsanov_theorem} \nonumber~.
\end{equation}

These probability measures $\mathbb{P}$ and $\mathbb{Q}$ are related by the Radon-Nikodym derivative, which is defined as follows, 
\begin{equation}
    \frac{d \mathbb{Q}}{d \mathbb{P}} = \exp\left( 
    - \int_0^t \gamma_s dW_s - \frac{1}{2} \int_0^t \gamma_s^2 ds
    \right) ~. \nonumber \label{eq:Camertn_Martin_Girsanov}
\end{equation} \\

Using the CMG theorem, we can estimate statistical quantities under a more general probability measure $\mathbb{Q}$. 
%Moreover, the CMG theorem can be used for the maximum likelihood estimation problem. 
Since the CMG theorem provides explicit transformation of probability measures, 
the maximization of the likelihood ratio can be a substitution of the maximum likelihood, 
\begin{equation}
    \begin{split}
    \max_{\theta} \mathbb{Q}_\theta(A) &= \max_{\theta} \int_A \frac{d \mathbb{Q}_\theta}{d \mathbb{P}}(x) ~d \mathbb{P} (x) \\
    &\leq 
    \int_A  \max_{\theta} \left\{\frac{d \mathbb{Q}_\theta}{d \mathbb{P}}(x) \right\} ~d \mathbb{P} (x) ~.
    \end{split} \nonumber
\end{equation}
Thus, we can estimate the most probable parameters by maximizing the likelihood ratio. 

Now, we can apply the CMG theorem to the inference problem of the OU model. The CMG theorem lets the SDE \eqref{eq:Ornstein_Uhlenbeck_main} transform into the following 
\begin{equation}
    d\tilde{\boldsymbol{X}}_t = - \tilde{\boldsymbol{\gamma}_t} dt + d\boldsymbol{W}_t \nonumber
\end{equation}
where $\tilde{\boldsymbol{X}}_t = \Sigma^{-1/2}\boldsymbol{X}_t$ and $\tilde{\boldsymbol{\gamma}_t} =- \Sigma^{-1/2} \mathrm{J}\boldsymbol{X}_t$. 
More general transformation can be done by the \emph{Lamperti transformation} that provides a systematic variable transformation rule so that a given SDE with multiplicative noise transforms to another SDE with an additive noise \cite{iacus2008simulation}.

Therefore, the likelihood ratio of the OU model becomes as follows, 
\begin{equation}
    \begin{split}
          \frac{d \mathbb{Q}_{\mathrm{J}}}{d \mathbb{P}} 
          &= \exp\left( 
    - \int_0^t \tilde{\boldsymbol{\gamma}}_s^\top d\tilde{\boldsymbol{X}}_s - \frac{1}{2} \int_0^t \tilde{\boldsymbol{\gamma}}_s^\top\tilde{\boldsymbol{\gamma}}_s ds
    \right) \\
          &= \exp\Big( 
     \int_0^t (\mathrm{J}\boldsymbol{X}_s)^\top \Sigma^{-1} d\boldsymbol{X}_s \\
     &\qquad\qquad - \frac{1}{2} \int_0^t (\mathrm{J}\boldsymbol{X}_s)^\top \Sigma^{-1} (\mathrm{J}\boldsymbol{X}_s)ds
    \Big) ~,
    \label{eq:CMG_theorem_OU}
    \end{split}
\end{equation}
where we used the symmetry of the covariance matrix and definition of the square matrix, $\Sigma^{1/2}\Sigma^{1/2} = \Sigma$. \\

Since the likelihood ratio \eqref{eq:CMG_theorem_OU} is a convex function of the coupling matrix, its derivative with respect to the coupling matrix gives the equation to solve the maximum likelihood estimator. 
So the derivative of the likelihood ratio is 
\begin{equation}
    \begin{split}
        \frac{\partial}{\partial \mathrm{J}} \log \frac{d \mathbb{Q}_{\mathrm{J}}}{d \mathbb{P}} 
    &= - \int_0^t \Sigma^{-1} d\boldsymbol{X}_s \boldsymbol{X}_s^\top
    - \int_0^t \Sigma^{-1} \mathrm{J}\boldsymbol{X}_s \boldsymbol{X}_s^\top ds \\
    &\xrightarrow{\mathrm{J}\to \mathrm{J}^*} ~ \mathrm{0} ~.
    \end{split} \nonumber
\end{equation}
This immediately leads the maximum likelihood ratio estimator
\begin{equation}
    \hat{\mathrm{J}} = \left(\int_0^t d\boldsymbol{X}_s \boldsymbol{X}_s^\top \right)
    \left(\int_0^t \boldsymbol{X}_s \boldsymbol{X}_s^\top ds \right)^{-1} ~. \nonumber \label{eq:MaxLikelihoodRatio_OU_model}
\end{equation}
To derive this solution, we used the fact that the inverse of the covariance is independent from the time and stochastic process.

The important consequence is that the maximum likelihood ratio based on the CMG theorem gives exactly the same solution as in the case of the path-likelihood maximization shown in \eqref{eq:MPLE_OU}.

\subsection*{Another derivation of optimal Wright-Fisher selection coefficients via Cameron-Martin-Girsanov theorem}
In this section, we will rederive the maximum path likelihood solution of the selection in the WF model using the CMG theorem.

We can write the Langevin equation for the Wright-Fisher diffusion as
\begin{equation}
    d\boldsymbol{X}_t = \left( 
\mathrm{C}(\boldsymbol{X}_t)\boldsymbol{s} + \boldsymbol{\mu}(\boldsymbol{X}_t)
\right) + \sqrt{\mathrm{C}(\boldsymbol{X}_t)}d \boldsymbol{W}_t ~. \nonumber
\end{equation}
Applying the formulation of the Radon-Nikodym derivative to this Langevin equation, we obtain 
\begin{widetext}
\begin{equation}
    \frac{{\rm d}\mathbb{Q}}{{\rm d}\mathbb{P}} = 
\exp\left(
\int_0^t (\mathrm{C}(\boldsymbol{x}_s)\boldsymbol{s}+\boldsymbol{\mu}(\boldsymbol{x}_s))^\top
 {\mathrm{C}(\boldsymbol{x}_s)}^{-1} d\boldsymbol{x}_s 
- \frac{1}{2} \int_0^t 
(\mathrm{C}(\boldsymbol{x}_s)\boldsymbol{s}+\boldsymbol{\mu}(\boldsymbol{x}_s) )^\top 
{\mathrm{C}(\boldsymbol{x}_s)}^{-1}
(\mathrm{C}(\boldsymbol{x}_s)\boldsymbol{s}+\boldsymbol{\mu}(\boldsymbol{x}_s))
ds
\right)\,.\nonumber
\end{equation} 
\end{widetext}
Since the logarithm of the RN derivative is a convex function, its derivative gives the solution that maximizes the RN derivative, 
\begin{equation}
\begin{split}
\frac{\partial}{\partial \boldsymbol{s}} 
\log \frac{{\rm d}\mathbb{Q}}{{\rm d}\mathbb{P}} &= 
\int_0^t d\boldsymbol{x}_s 
- \int_0^t 
(\mathrm{C}(\boldsymbol{x}_s)\boldsymbol{s}+\boldsymbol{\mu}(\boldsymbol{x}_s) )
ds \\
&\xrightarrow{\boldsymbol{s}\to \boldsymbol{s}^*} ~ \boldsymbol{0} ~.
\end{split} \nonumber
\end{equation}

Therefore, the solution equivalent to the maximum path-likelihood solution is obtained.

\begin{equation}
    \hat{\boldsymbol{s}} = 
\left(
\int_0^t 
\mathrm{C}(\boldsymbol{x}_s)ds
\right)^{-1}
\int_0^t (
d\boldsymbol{x}_s - \boldsymbol{\mu}(\boldsymbol{x}_s) ds
) \nonumber
\end{equation}

%--------- MPL for WF ---------%
\newpage
\subsection*{Effect of regularization strength  $\gamma$}\label{sup:regularization_dependency}
We report the influence of the regularization on the precision of the selection coefficients based on positive predictive value (PPV) curves. In this test, we chose the following different regularization values $\gamma \in \{10^{-3}, 0.1, 1, 5, 10 \} $. Through the all tests, we fixed the sampling interval as $\Delta t = 75$. For the other parameters, we use the same parameters that are used in the main section.

\begin{figure*}
\centering
\includegraphics[width=11.4cm]{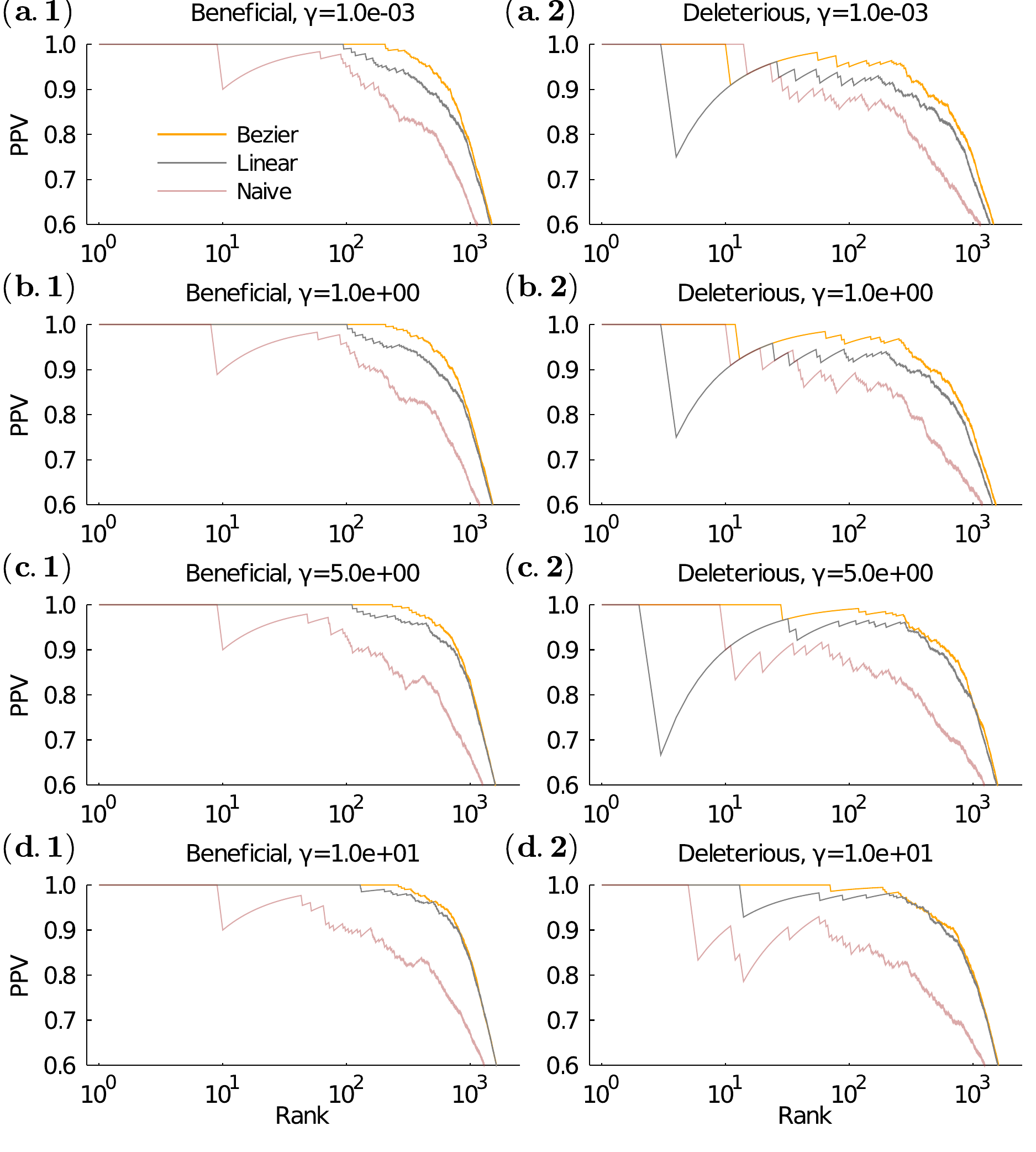}
\caption{\protect\rule{0ex}{5ex} 
 \textbf{Across a wide range of regularization values, B\'ezier interpolation achieves more accurate inference than linear interpolation.}
\textbf{(a.1)} PPV curves for beneficial selection coefficients using $\gamma=10^{-3}$. Other conditions are the same as in the main text. \textbf{(a.2)} PPV curves for deleterious coefficients using $\gamma=10^{-3}$. \textbf{(b)}, \textbf{(c)}, \textbf{(d)} the same type of figures but using $\gamma=1.0, \gamma=5.0$ and $\gamma=10.0$. MPL with linear interpolation is sensitive to the regularization strength, and larger regularization is necessary to make more precise inferences. However, the most accurate PPV using MPL with linear interpolation ($\gamma=30.0$) has almost the same performance as the least accurate inferences using MPL with B\'ezier interpolation (see \textbf{Supplementary Fig.~\ref{fig:compare_Bezier_linear_PPV}}). Moreover, the larger regularization induces a strong estimation bias, as shown in \textbf{Supplementary Fig.~\ref{fig:selections_storng_weak_reg}}.
\label{fig:PPV_regularization_dependency}}
\end{figure*}

\textbf{Supplementary Fig.~\ref{fig:PPV_regularization_dependency}} shows how inference accuracy depends on the regularization strength for MPL using different interpolation methods: piece-wise constant, linear, and B{\'e}zier interpolation.

In the case of the small to medium regularization values ($\gamma = 10^{-3}, 0.1, 1$), PPV curves using the B{\'e}zier interpolation are significantly higher than the PPV curves using other interpolation methods. As the regularization value increases, the difference between the PPV curves for linear and B{\'e}zier interpolations becomes smaller.

\begin{figure*}
\centering
\includegraphics[width=11.4cm]{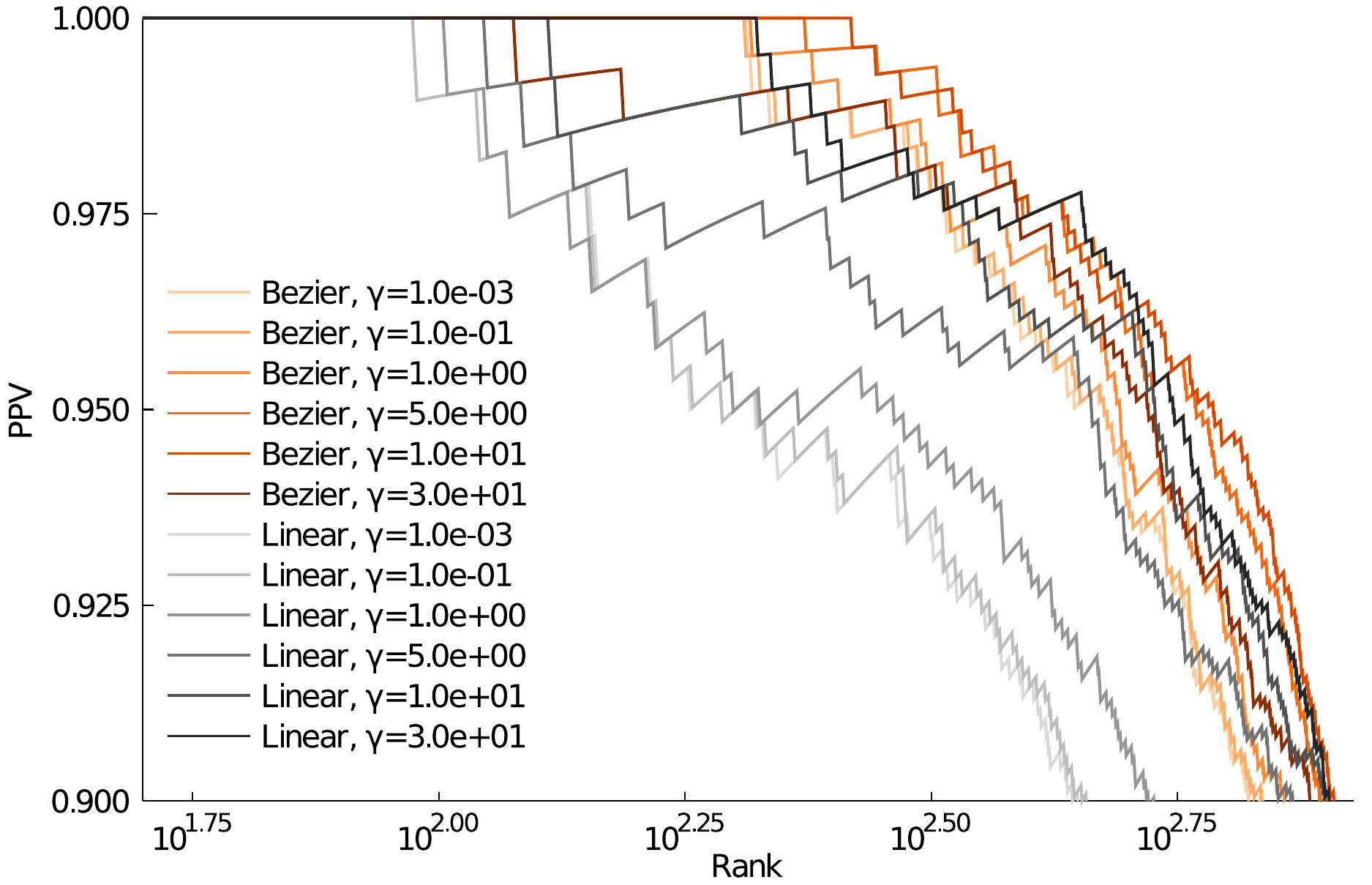}
\caption{\protect\rule{0ex}{5ex} 
\textbf{In a wide range of regularization, MPL with B\'ezier interpolation achieves higher PPVs than MPL with linear interpolation.} 
Here we show PPV curves between rank 60 and 900, where changes due to the different regularization values $\gamma=10^{-3}, 0.1, 1.0, 5.0, 10.0$ and $30.0$ are most noticeable. 
PPV curves of MPL with B\'ezier interpolation maintain high values stably. In contrast, PPV curves of MPL with linear interpolation are sensitive to the choice of the regularization strength and tend to be lower than those for MPL with B\'ezier interpolation. In the linear interpolation case, larger regularization yields higher the PPV curves, but also larger estimation bias.
\label{fig:compare_Bezier_linear_PPV}}
\end{figure*}

\textbf{Supplementary Fig.~\ref{fig:compare_Bezier_linear_PPV}} shows that MPL with B{\'e}zier interpolation outperforms MPL with linear interpolation for any regularization strength $\gamma$. The best PPV curves of MPL with linear interpolation are still lower than the majority of PPV curves for MPL using B\'ezier interpolation. Moreover, although a large regularization improves the PPV curves of MPL with linear interpolation, due to the strong regularization effect, the estimated selection coefficients are strongly biased and are underestimated as shown in \textbf{Supplementary Fig.~\ref{fig:selections_storng_weak_reg}}.

\begin{figure*}
\centering
\includegraphics[width=11.4cm]{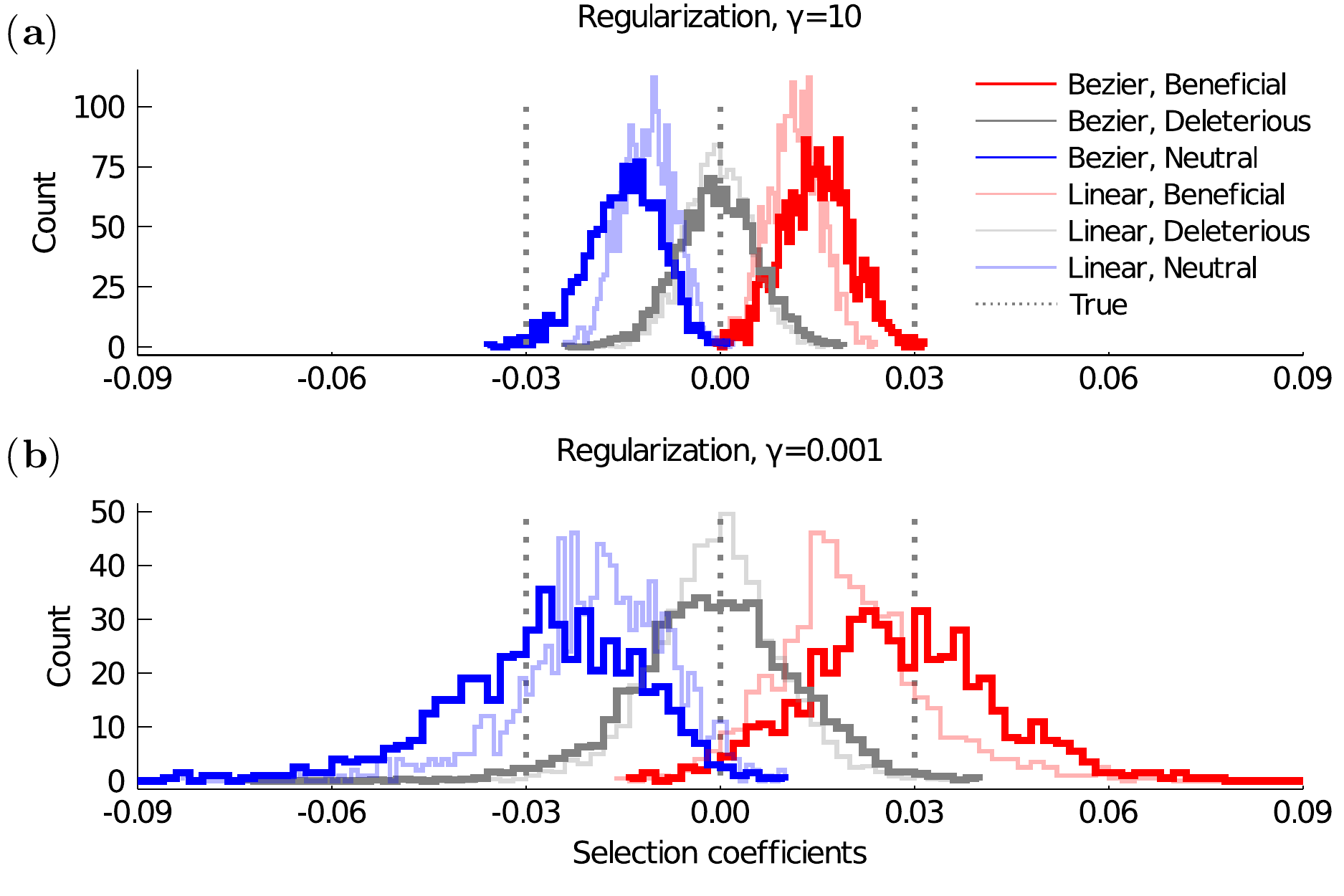}
\caption{ 
\textbf{MPL with B\'ezier interpolation reduces estimation bias in a wide range of regularization, and small regularization is needed to avoid strong estimation bias.} 
 \textbf{(a)} Distribution of inferred selection coefficients using a strong regularization $\gamma=10$. Other conditions are the same as in the main text. \textbf{(b)} Estimated selection coefficients using a weak regularization $\gamma=10^{-3}$. 
Smaller regularization $\gamma=10^{-3}$ reduces estimation bias, especially for MPL with linear interpolation.
\label{fig:selections_storng_weak_reg}}
\end{figure*}

\newpage
\subsection*{Effect of sampling interval $\Delta t$ }\label{sup:sampling_time_dependency}
Here, we discuss the effects of the sampling interval $\Delta t$ on the different interpolation methods in detail. 
In this study, the model parameters for the population size and mutation rate are the same as in \hyperref[{sec:wf_bezier}]{the main text}, and the regularization coefficient is fixed as $\gamma = 0.1$.

\begin{figure*}
\centering
\includegraphics[width=11.4cm]{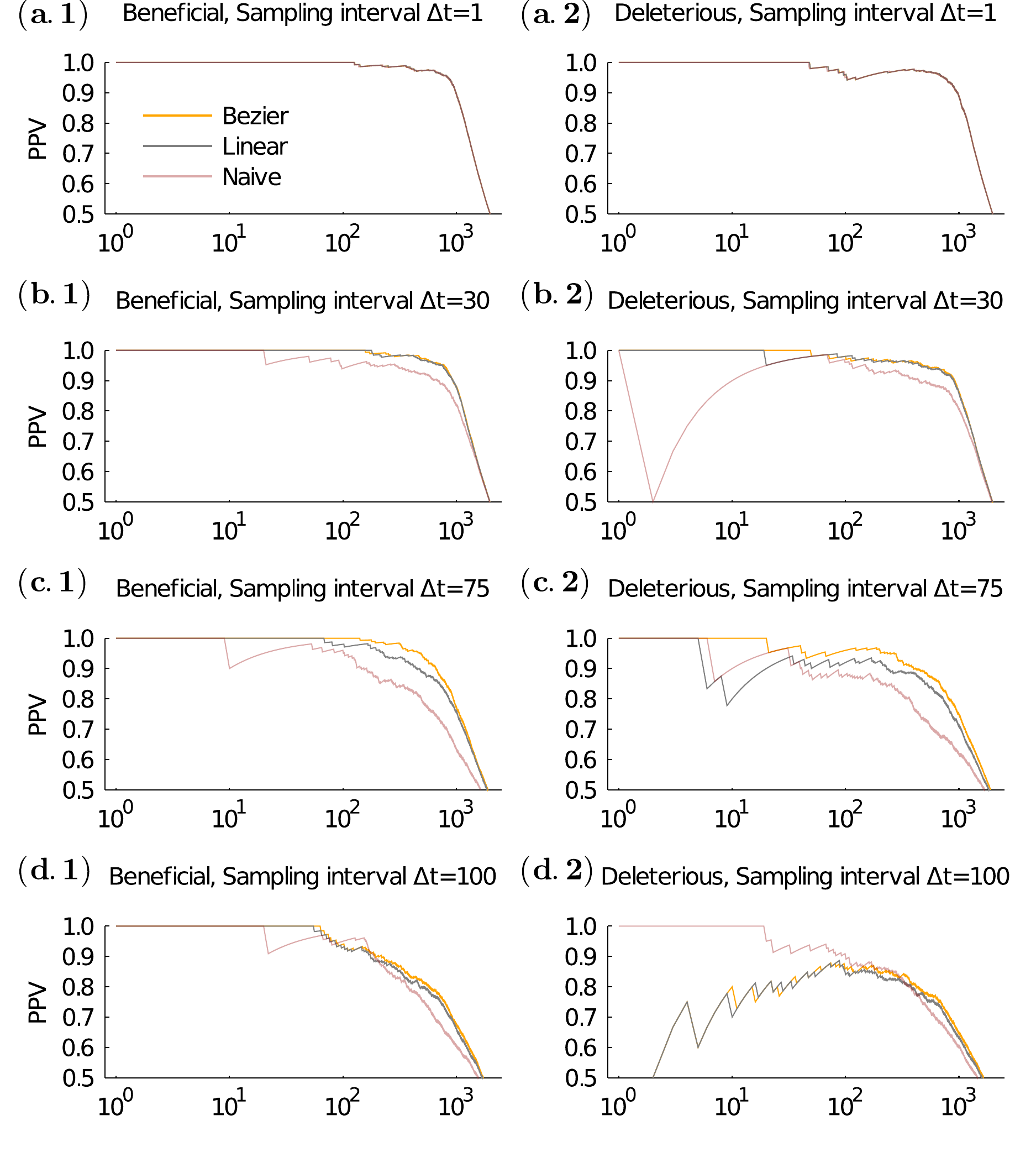}
\caption{
%\protect\rule{0ex}{5ex} 
\textbf{As the sampling interval increases, the advantage of the B\'ezier interpolation becomes more notable.}
\textbf{(a.1)} PPV curves for beneficial selection coefficient inference using the sampling interval $\Delta t=1$, using MPL with B\'ezier, linear, and piece-wise constant interpolations. Other conditions are the same as in the main text ($\gamma=0.1, N=10^3$, and $\mu=10^{-3}$) \textbf{(a.2)} PPV curves for deleterious selection coefficient inference using $\Delta t=1$. Subplots, \textbf{(b)}, \textbf{(c)} and \textbf{(d)} are the same type of figures but for $\Delta t = 30, 75$, and $100$, respectively. As the sampling interval increases, the inference accuracy decreases in the PPV sense. However, inferences using B\'ezier interpolation degrade more slowly than other methods. For the longest sampling interval ($\Delta t = 100$), consecutive time points are poorly correlated. As a result, none of the interpolation methods can completely accurately infer selection coefficients, and hence the PPV curves roughly converge.
\label{fig:ppv_sampling_time_dependency}}
\end{figure*}

\textbf{Supplementary Fig.~\ref{fig:ppv_sampling_time_dependency}} shows PPV curves for estimated selection coefficients using MPL with piece-wise constant, linear, and B{\'e}zier interpolation depending on various sampling intervals $\Delta t \in \{1, 10, 30, 75, 100 \}$. 
% including the figure for MPL and SL

For $\Delta t = 1, 10$, there is no difference among these methods. 
However, when $\Delta t=30$, the PPV curves for the piecewise constant case deteriorate compared with the other methods and the ones for the linear and B\'ezier interpolations are indistinguishable. This is consistent with \hyperref[{subsec:recovery_of_decaying_correlation}]{the argument} in the main section: the characteristic time scale, $\gamma\Delta t$, is not so large that nonlinear effects are noticeable, hence PPV curves for the linear and B\'ezier interpolation are indistinguishable.

In the $\Delta t = 75$ case, the PPV curves of the MPL with B\'ezier interpolation are systematically higher than the cases of MPL with linear interpolation, hence MPL with B\'ezier interpolation outperforms other approaches.

In general, as the time interval increases,  B\'ezier interpolation has a greater advantage in capturing the underlying dynamics of trajectories (\textbf{Supplementary Fig.~\ref{fig:ppv_sampling_time_dependency}}). However, for large enough time gaps, all interpolation methods suffer because data is sampled too sparsely to reveal any information about the underlying dynamics. For large enough $\Delta t$, there is no connection between the covariances at consecutively sampled points, and ``trajectory information'' is no longer contained in the data. This is also consistent with the negligible size of the autocorrelation for off-diagonal covariances at very large time gaps.

\newpage
\subsection*{Positive semidefiniteness of the interpolated covariance}\label{sup:positive_semidefinitness}
The eigenvalues of the covariance matrix are strictly non-negative. This positive semi-definiteness is an essential property of the covariance matrix and is practically important. We numerically confirmed the positive semidefiniteness of interpolated covariance matrices using the B\'ezier interpolation.

To evaluate the positive semidefiniteness, we generated a test data set by running the WF model 100 times. The dependent parameters of the WF model are the same as \hyperref[{sec:wf_bezier}]{the main text}. Then, we estimated integrated covariance matrices and their covariance matrix eigenvalues for different interpolation methods and different sampling intervals. 

\begin{figure*}
\centering
\includegraphics[width=0.5\linewidth]{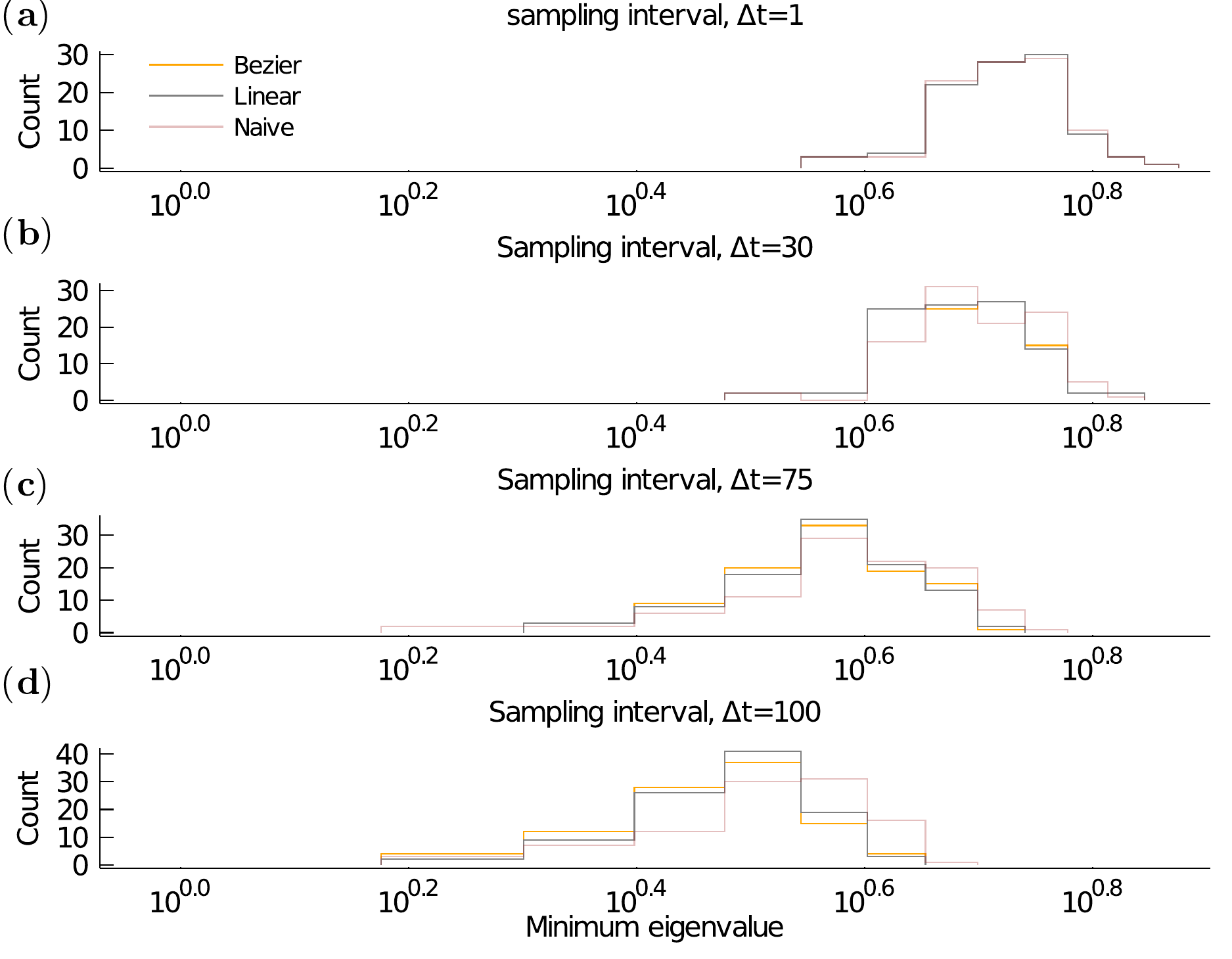}
\caption{\textbf{Covariance matrix with B\'ezier interpolation maintains positive semidefiniteness.}
Comparison of the minimum eigenvalue distributions of the integrated covariance matrices: As the $\Delta t$ increases, the minimum eigenvalues are smaller, but they remain nonnegative values. Thus, all the integrated covariances are positive definite.
\label{fig:minimu_eigenvalues}}
\end{figure*}

In either interpolation method, the eigenvalue distribution of the integrated covariance matrix showed little change, and only positive eigenvalues were observed in each case (\textbf{Supplementary Fig.~\ref{fig:minimu_eigenvalues}}).

%\clearpage

\newpage
\subsection*{Selection coefficient inference without off-diagonals of integrated covariance elements}\label{sup:MPL_SL_comparison}
As shown in \hyperref[{subsec:recovery_of_decaying_correlation}]{the main text}, B\'ezier interpolation is better than linear interpolation in the sense of the more accurate reconstruction of the covariance matrix depending on perfectly observed trajectories (when the sampling interval $\Delta t = 1$) from the covariance matrix depending on ``sparsely" observed trajectories, especially for the ``off-diagonal" elements (corresponding to pairwise covariances $C_{ij}=x_{ij}-x_ix_j$, with $i\neq j$) of the integrated covariance matrix. On the other hand, the difference between linear and B\'ezier interpolation for the ``diagonal" elements (variance $C_{ii} = x_i(1-x_i)$) was relatively minor. To understand how exactly this observation is associated with the accuracy of the selection coefficients, we examine the effect of the off-diagonal entries of the integrated covariance matrix on the selection coefficients in this section.

\begin{figure*}
\centering
\includegraphics[width=11.4cm]{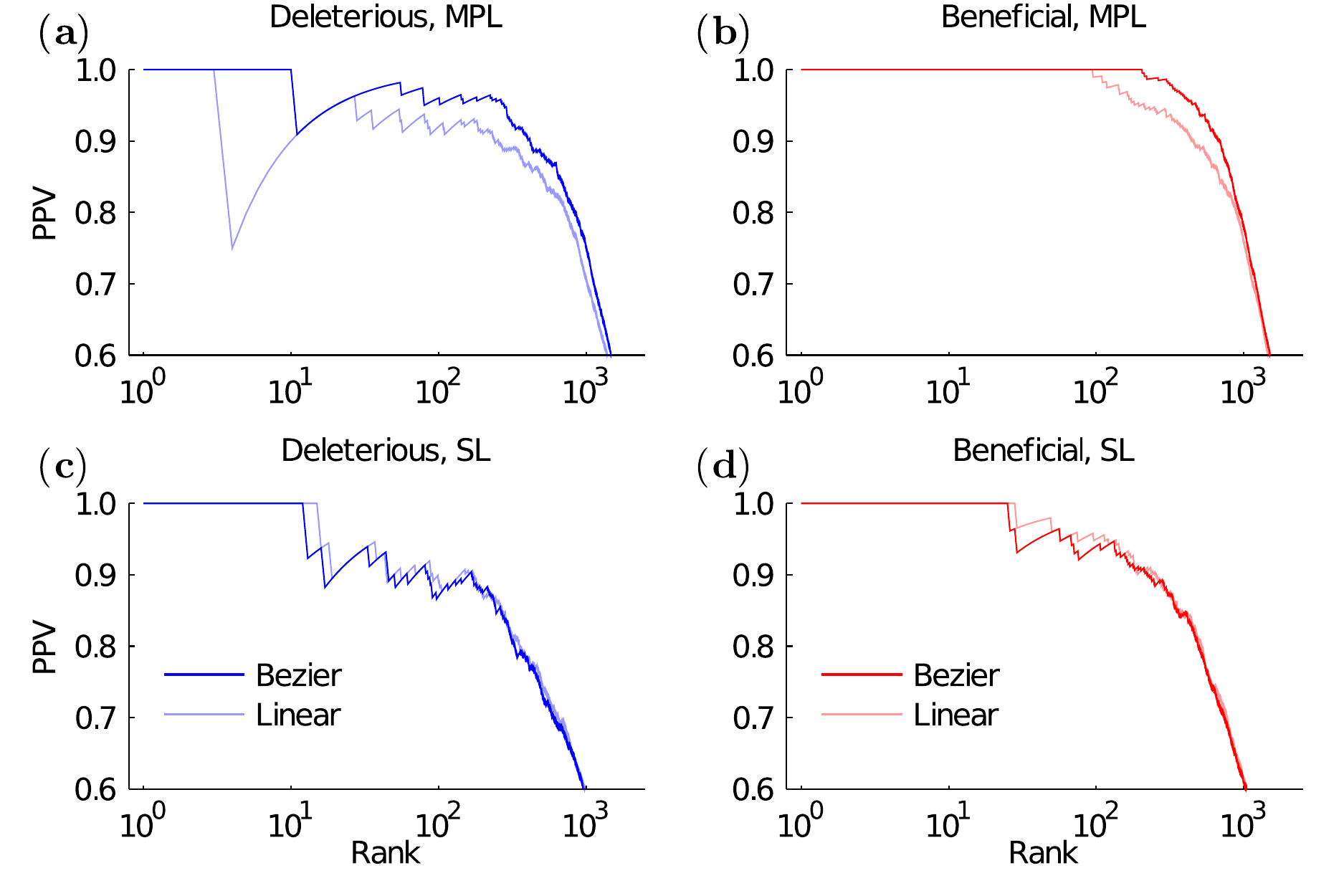}
\caption{ \textbf{Improvement of the selection inference accuracy is due to the accurate restoration of pairwise frequencies.}
PPV curves for \textbf{(a)} deleterious and \textbf{(b)} beneficial selection coefficients using MPL methods. The sampling interval is fixed as $\Delta t = 75$.  PPV curves for \textbf{(c)} deleterious and \textbf{(d)} beneficial selection coefficients, but using the single locus (SL) method, a simplified version of MPL which sets off-diagonal elements of the covariance matrix to zero. B\'ezier interpolation improves the precision of MPL, but the choice between linear and B\'ezier interpolation does not significantly affect the accuracy of SL. This implies that the accurate estimation of pairwise frequencies (corresponding to off-diagonal covariances) improves selection inference accuracy.
\label{fig:PPV_SL_MPL}}
\end{figure*}

\textbf{Supplementary Fig.~\ref{fig:PPV_SL_MPL}} shows the inference accuracy for both deleterious and beneficial mutations using MPL and the single locus (SL) method, a simplified inference method that ignores the off-diagonal of the integrated covariance matrix. 

The PPV of MPL with B\'ezier interpolation achieves systematically higher values than the PPV of MPL with linear interpolation. However, the difference between linear and B\'ezier interpolation becomes unclear for inferences using SL. 
Thus, the main reason MPL with B\'ezier interpolation can infer better than MPL with linear interpolation is the accurate estimation of off-diagonal covariances (including pairwise frequencies).

\newpage
%--------- For OU processes ---------%
\subsection*{Ornstein-Uhlenbeck process inference comparison}
In this section, we report a more detailed analysis of the estimated coupling parameters of OU processes. The input data sets for the inference are the same as in \hyperref[{sec:inverse_Ornstein_Uhlenbeck}]{ the main section}. To compare the inference accuracy between various inference methods, besides the path-likelihood-based methods, we included mean-field theory-based inference. In this approach, the effective solution is given by the inverse of the integrated covariance matrix, which effectively predicts interaction matrices for input data following an equilibrium distribution\cite{morcos2011direct}.

\begin{figure*}
\centering
\includegraphics[width=11.4cm]{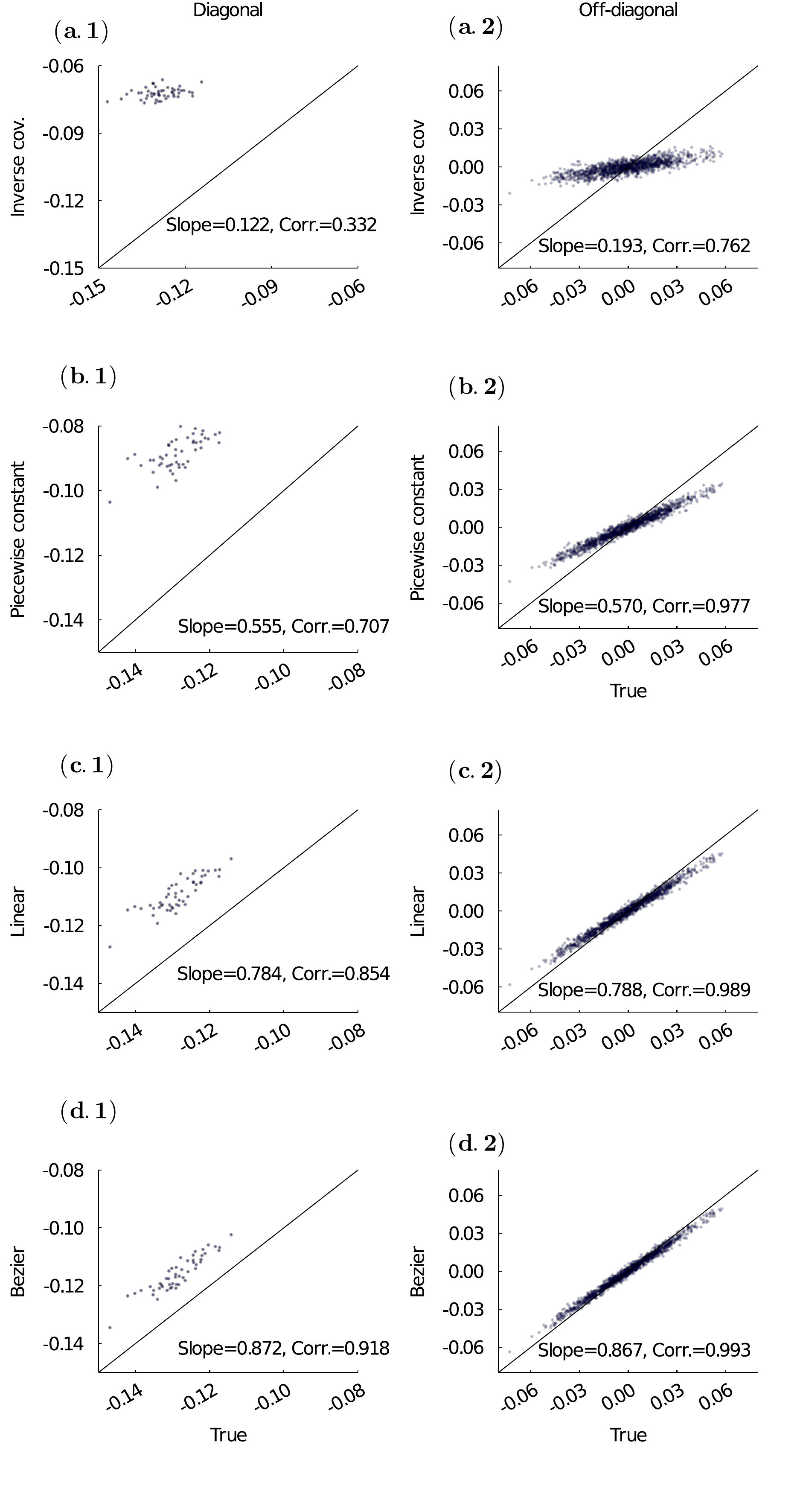}
\caption{ \textbf{Path-likelihood-based inference method with B\'ezier interpolation achieves the best inference accuracy. }
Comparison of true and inferred OU process interaction matrices. Mean-field based methods were used for \textbf{(a.1)} diagonal and (\textbf{a.2)} off-diagonal elements of the interaction matrix. Panels \textbf{(b)}, \textbf{(c)}, and \textbf{(d)} are the same type of plots, but using the path-likelihood-based inference with piecewise constant, linear, and B\'ezier interpolation, respectively. Among all the methods, inference with B\'ezier interpolation achieves the highest accuracy in terms of Pearson's correlation coefficient and regression slope value.
\label{fig:scatter_Bezier_linear}}
\end{figure*}

\textbf{Supplementary Fig.~\ref{fig:scatter_Bezier_linear}} shows comparisons of a true interaction matrix and estimated interactions. The accuracy of the path-likelihood-based methods is significantly better than the the inverse of the covariance matrix in terms of Pearson's correlation and linear regression's slope. This is an anticipated result since the input data sets were generated from the relaxation processes, and the probability distributions that characterize these dynamics are in non-steady states. Therefore, MPL methods outperform inference methods assuming equilibrium states. 

The path-likelihood-based inference method with B\'ezier interpolation achieves the best inference accuracy for both diagonal and off-diagonal interaction matrix elements in terms of Pearson's correlation coefficients and regression slope values. 

\begin{figure*}
\centering
\includegraphics[width=11.4cm]{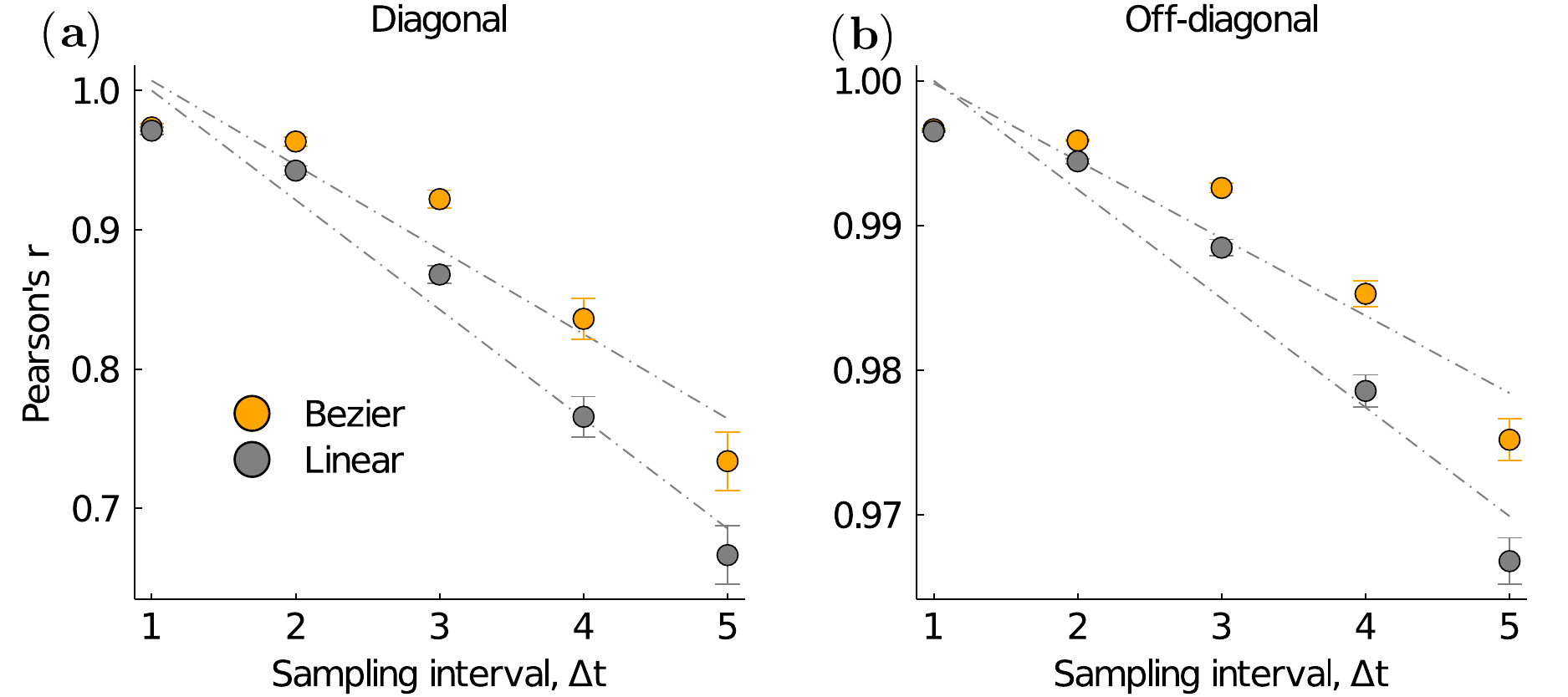}
\caption{ 
\textbf{As the sampling interval increases, the advantage of B\'ezier interpolation becomes more pronounced.}
Dependence of Pearson's correlation coefficients for \textbf{(a)} diagonal and \textbf{(b)} off-diagonal interaction matrices on the sampling interval.
Pearson's correlation coefficients for  B\'ezier interpolation are significantly higher than the ones for linear interpolation, especially when the sampling interval is large.
\label{fig:Pearson_dt_dependency}}
\end{figure*}

\textbf{Supplementary Fig.~\ref{fig:Pearson_dt_dependency}} shows sampling interval dependence for Pearson's correlation coefficients between true interaction matrices and inferred interaction matrices. The input data sets and conditions of the inferences are the same as the main text. As the sampling interval regime increases, the difference between Pearson's $r$ of linear and B\'ezier interpolations becomes more pronounced, and the inferences using B\'ezier interpolation achieve higher Pearson's $r$ values among all sampling intervals.

%\clearpage

%\clearpage

\section*{References}
\newbibstartnumber{46} % This should be the number of references in the main text + 1
\bibliographystyle{naturemag}
\bibliography{Bezier_Interpolation}

%\clearpage
%\begin{figure*}[t]
%\includegraphics[width=\textwidth]{figures/figs2-performance.pdf}
%\caption{
%\textbf{Supplementary figure title.} 
%And its caption. 
%\label{fig:supplementary_figure}}
%\end{figure*}

%\newpage
%\include{supplement}

%%%%%%%%%%%%%%%%%%%%%%%%%%%%%
% WORD COUNT                %
%%%%%%%%%%%%%%%%%%%%%%%%%%%%%
%TC:ignore
%\newpage
%\section*{Word Counts}
%This section is \textit{not} included in the word count. 

%\subsection*{Statistics on word count}
%\detailtexcount

%TC:endignore
%the command above ignores this section for word count

\end{document}